\documentclass[epj]{svjour}
\usepackage{graphics}
\usepackage[dvips]{color}

\def\be{\begin{equation}}
\def\ee{\end{equation}}
\newcommand{\dil}{{\rm dil}}
\newcommand{\reg}{{\rm reg}}
\newcommand{\conf}{{\rm conf}}

\begin{document}

\title{
The $2D$ $XY$ model on a finite lattice with structural disorder:
\newline quasi-long-range ordering under realistic conditions}

\titlerunning{The $2D$ $XY$ model on a finite lattice with structural disorder}

\author{O. Kapikranian\inst{1,2}, B. Berche\inst{1} \and Yu. Holovatch\inst{2,3}}                     

\authorrunning{O. Kapikranian, B. Berche and Yu. Holovatch}

\institute{Laboratoire de Physique des Mat\'eriaux,
    Universit\'e Henri Poincar\'e,  Nancy 1,\\
    F-54506 Vand\oe uvre les Nancy Cedex, France
    \and
    Institute for Condensed Matter Physics,
    National Academy of Sciences of Ukraine, \\
    Lviv, 79011 Ukraine
    \and
    Institut f\"{u}r Theoretitsche Physik, Johannes Kepler Universit\"at,
    Linz, 4040 Austria}

\date{Received: date / Revised version: date}
%
\abstract{ We present an analytic approach to study
concurrent influence of quenched non-magnetic site-dilution and
finiteness of the lattice on the 2D XY model. Two significant
deeply connected features of this spin model are: a special type
of ordering (quasi-long-range order) below a certain temperature
and a size-dependent mean value of magnetisation in the
low-temperature phase that goes to zero (according to the
Mermin-Wagner-Hohenberg theorem) in the thermodynamic limit. We
focus our attention on the asymptotic behaviour of the spin-spin
correlation function and the probability distribution of
magnetisation. The analytic approach is based on the spin-wave
approximation valid for the low-temperature regime and an
expansion in the parameters which characterise the deviation from
completely homogeneous configuration of impurities. We further
support the analytic considerations by  Monte Carlo simulations
performed for different concentrations of impurities and compare
analytic and MC results. We present as the main quantitative
result of the work the exponent of the spin-spin correlation
function power law decay. It is non universal depending not only
on temperature as in the pure model but also on concentration of
magnetic sites. This exponent characterises also the vanishing of
magnetisation with increasing lattice size.
\keywords{$XY$ model
-- topological transition  -- random systems.}
\PACS{{05.50.+q}{Lattice theory and statistics (Ising, Potts,
etc.)} \and
      {64.60.Fr}{Equilibrium properties near critical points,
        critical exponents} \and
      {75.10.Hk}{Classical spin models}
     } 
} 
\maketitle

\section{Introduction}
\label{intro}

The quasi-long-range ordering (QLRO) is a special feature of a
number of important many-particle systems including
two-dimensional solids, magnets, Bose fluids, liquid crystals
\cite{Chaikin95,Nelson02,MikeskaSchmidt70}.
As it is known by now the 2D $XY$ model
serves as an archetype capturing special features of QLRO in these
systems. Here we will focus on this particular model keeping in
mind that the results can be generalised for some other similar
models. The regular model, described by the Hamiltonian

\begin{equation}
H_{\reg} = -\frac{1}{2}\sum_{\bf r}\sum_{\bf r'}J({\bf
r-r'})\left(S^x_{\bf r}S^x_{\bf r'}+S^y_{\bf r}S^y_{\bf
r'}\right)\ ,\label{Hpure}
\end{equation}
has been investigated in great detail, and although most of its properties
are known, no exact
solution was found. In (\ref{Hpure}) ${\bf r}$ and ${\bf r'}$ span
sites of a two-dimensional lattice, $J({\bf r})$ is the nearest
neighbours interaction potential, $S^x_{\bf r}$, $S^y_{\bf r}$ are
the components of a classical ``spin" ${\bf S}_{\bf r}$, the
coefficient 1/2 stands to prevent double count of each bond. The
spin-wave approximation (SWA) applied by F. Wegner
~\cite{Wegner67} to analyse the $2D$ $XY$ model leads to a result
very close to recent Monte Carlo computations in the region of
low enough temperatures. In particular, the presence of a special type
of ordering -- the QLRO -- manifests itself in the
power law decay of the spin-spin correlation function:

\begin{equation}\label{CorrFuncPure}
\left<{\bf S_{r}S_{r+R}}\right>\sim R^{-\eta^{\reg}}\ ,
\end{equation}
where $R$ is the distance between the spins. The exponent $\eta^{\reg}$ given by the
SWA is non-universal:

\begin{equation}\label{EtaReg}
\eta^{\reg} = kT/(2\pi J)\ .
\end{equation}

The detailed description of properties of the model given by
V.L. Berezinskii \cite{Berezinskii71},
and J. M. Kosterlitz and D. J. Thouless (BKT)
\cite{KosterlitzThouless73,Kosterlitz74}
is based on the hypothesis that certain local spin
configurations, named topological defects, are responsible for
the QLRO and the behaviour of the system near the transition to
the QLRO phase (the BKT transition) at the temperature $T_{BKT}$.
The intuitive analogy with the
transition in electrolytes was used in those works. A further
analytical basement for this approach can be found in the work of
J. Villain~\cite{Villain75}.

There are two important aspects which differ the ideal 2D $XY$ model
from the systems that can be met in nature: real physical systems
{\em are always finite} and {\em possess structural defects}. The
finite size effect already has been widely explored in the pure
(undiluted) $XY$ model. The interest of this question is that in any
2D $XY$ system of finite size, magnetisation is non-vanishing
\cite{BramwellHoldsworth93} and goes to zero only when the lattice
becomes infinite as it should be according to the
Mermin-Wagner-Hohenberg theorem \cite{MerminWagner66,Hohenberg67}.

The decay of the magnetisation has a power law
form (below the BKT transition temperature)
that can be easily found in the SWA ~\cite{TobochnikChester79,Archambault97}:
\begin{equation}\label{MagnDecay}
\left<m\right>\sim N^{-\frac{\eta^{\reg}}{4}}\ ,
\end{equation}
with the same $\eta^{\reg}$, Eq.(\ref{EtaReg}), that stands in the
correlation function (\ref{CorrFuncPure}). Here, $N$ is the total number of spins.
The same result can be obtained also from
the finite size scaling (see e.g. Ref.~\cite{BercheParedes05} in a similar
context.).

The recent works of S. T. Bramwell et al.
\cite{Archambault97,BramwellEtAl01,BanksBramwell05}
give deep analysis of
the magnetisation probability distribution
which they claim to be non-Gaussian and of universal form,
independent of both system size and critical exponent
$\eta^{\reg}$.

Structural disorder as site- or bond-dilution deserves much attention
since it moves an ideal model closer towards true physical systems
which can be found in nature. However the number of works dedicated to
this aspect in the 2D $XY$ model is
not mirrored in the great importance of the topic. Harris
criterion \cite{Harris74} implies that energy-coupled disorder has no
effect on the universal properties (e.g. the critical exponents) {\em at
the transition temperature}. The BKT universality class (and in
particular the celebrated $\eta(T_{BKT})=\frac 14$) is thus unchanged
by the introduction of quenched disorder, but
one can expect highly non-trivial dependence of the low-temperature
characteristics, like the spin-spin correlation function,
on the concentration of
spin-vacancies~\cite{BercheEtAl02}.

A non-magnetic site can change the interaction between topological
defects which are responsible as it was mentioned above for the QLRO
~\cite{PaulaEtAl05,WysinEtAl05,LeonelEtAl03}.
The character of this influence is not completely clear up to now, for example,
the question: when the QLRO disappears as the concentration of vacant sites
increases, has got different answers~\cite{LeonelEtAl03,BercheEtAl02,SurunganOkabe05}.
As it appears now, the most convincing
scenario is that the QLRO remains up to concentrations very close to the percolation
threshold \cite{BercheEtAl02,SurunganOkabe05}. However we do
not touch this question focusing
mostly on the region far from the percolation
threshold.

In this paper we investigate the concurrent influence of quenched
site-dilution and finite size of the lattice on the properties of
the $2D$ $XY$ model. These two modifications together present a nice
approach to investigation of real physical systems. To quantify the
disorder-induced changes in the QLRO phase we pay attention to the
spin-spin correlation function exponent of the 2D $XY$ model with
quenched site-dilution, $\eta^{\dil}$. It describes not only the
decay of the correlation function with the distance but also the
vanishing of the magnetisation in a finite system with increase of
the lattice size and the divergence of the susceptibility in the
same limit. The analytic approach we use here relies on the SWA and
a perturbation expansion and is verified by MC simulations.

Also we will perform Monte Carlo simulations for a wide range of 2D $XY$-spin
systems of different sizes at different temperatures and with different
concentrations of impurities. Systems explored in computer experiments
are always finite, thus they possess a non-vanishing mean value of
magnetisation. The instantaneous magnetisation, which is the scalar value of the
total sum of the spins divided by the number of sites, measured in a given state
from the thermodynamical ensemble of states of the system is distributed
with a certain law. The form of this distribution is the point of our interest
as well.

The structure of the paper is the following:
In the second section we give a description of the model
and calculate analytically the spin-spin correlation
function combining the SWA and perturbation expansion, we support
the result by MC simulations.
In Section 3 more details about
MC simulations can be found. The results
are presented in the form of ring functions and probability
distribution functions of magnetisation. We analyse the plots and add an analytic
calculation of the moments of magnetisation.
We discuss the analytic and MC results and sketch the plans of future work
in Conclusions and give two appendices with technical details of the calculations.

\section{The spin-spin correlation function}
\label{sec1}

In this section we give description of the diluted 2D $XY$ model and
explain the expansion applied to analyse the
asymptotic behaviour of the spin-spin correlation function in the
low temperature limit. The comparison with our Monte Carlo results
is added to support the analytic approach.

\subsection{The model}
\label{subsec1}

The regular 2D $XY$ model (\ref{EqH_reg}) is equally described in
the angle variables $\theta_{\bf r}$'s that are the angles between
the spins and a certain fixed direction by the Hamiltonian

\begin{equation}
H_{\reg} = -\frac{1}{2}\sum_{\bf r}\sum_{\bf r'}J({\bf
r-r'})\cos(\theta_{\bf r}-\theta_{\bf r'})\ ,\label{EqH_reg}
\end{equation}
since $S^x_{\bf r}S^x_{\bf r'}+S^y_{\bf r}S^y_{\bf
r'}=\cos(\theta_{\bf r}-\theta_{\bf r'})$ for unit length spins.
All the notations are the same as in (\ref{Hpure}).

We define a set of occupation numbers $c_{\bf r}$'s that introduce
disorder into the lattice:

\begin{equation}
c_{\bf r} = \left\{ \begin{array}{ll}
1, & \textrm{if the site ${\bf r}$ has a spin;}\\
0, & \textrm{if the site ${\bf r}$ is empty.}
\end{array} \right.\label{OccNumb}
\end{equation}
The Hamiltonian modified with these numbers,

\begin{equation}
H = -\frac{1}{2}\sum_{\bf r}\sum_{\bf r'}J({\bf
r-r'})\cos(\theta_{\bf r}-\theta_{\bf r'}) c_{\bf r}c_{\bf r'}\ ,
\end{equation}
will describe the model on a lattice with dilution. Setting a
certain sequence of numbers $\{c_{\bf r}\}$ we are able to realize
any configuration of lattice dilution with any desirable
concentration of magnetic sites $c=\overline{c_{\bf r}}$. We are
interested in some thermodynamical quantities which are dependent
in this case on the configuration of impurities. To obtain
observable values we will average the quantities of interest over
all the  possible configurations of non-magnetic sites; this is
referred to as quenched disorder in the literature as in contrast to annealed
disorder when magnetic and non-magnetic sites are in equilibrium
and the configurational averaging has to be made already
in the partition function~\cite{Brout59}. We denote the
configurational averaging as $\overline{(...)}$:
\begin{equation}
\overline{(...)} = \prod_{\bf r}\sum_{c_{\bf
r}=0,1}[c\delta_{1-c_{\bf r},0}+(1-c)\delta_{c_{\bf
r},0}](...) \label{Eq-ConfAvg}
\end{equation}
($\delta_{i,j}$ are Kroneker deltas), here and below index ${\bf r}$
in sums and products spans all sites of a 2D square lattice.

Since we restrict ourselves to the low temperature
phase of the model we assume that the directions of spins on neighboring
sites do not differ essentially. This allows us to pass to the SWA
replacing $\cos\left(\theta_{\bf r}-\theta_{\bf r'}\right)$ in the
Hamiltonian with a quadratic form $1-\frac{1}{2}\left(\theta_{\bf
r}-\theta_{\bf r'}\right)^2$. All the main features of the model
are preserved in the low temperature limit in the Hamiltonian

\begin{equation}
H = H_0 + \frac{1}{4}\sum_{\bf r}\sum_{\bf r'}J({\bf
r-r'})\left(\theta_{\bf r}-\theta_{\bf r'}\right)^2c_{\bf r}c_{\bf
r'} \label{EqH_dil}
\end{equation}
where the first term in the expression can be regarded just as a
shift in the energy scale, from now on we denote the second term
by $H$ for simplicity.

Using Fourier transformation of the variables:
\begin{eqnarray}\label{ThetaFour}
\theta_{\bf r}&=&\frac{1}{\sqrt{N}}\sum_{\bf k}e^{i{\bf
kr}}\theta_{\bf k}, \quad \theta_{\bf
k}=\frac{1}{\sqrt{N}}\sum_{\bf r}e^{-i{\bf kr}}\theta_{\bf r},
\\\nonumber
J({\bf r})&=&\frac{1}{N}\sum_{\bf q}e^{i{\bf qr}}\nu({\bf q}),
\quad \nu({\bf q})=\sum_{\bf r}e^{-i{\bf qr}}J({\bf r}),
\end{eqnarray}
where ${\bf k}$ runs over the 1st Brillouin zone,
one arrives at:

\begin{equation}\label{Eq-HamDilute}
H\ =\ c^2 H_{\reg}\ +\ H_{\rho}\ +\ H_{\rho^2}\ ,
\end{equation}
with

\begin{eqnarray}
H_{\rho} &\equiv& -\ cJ\sum_{\bf k, k'}\gamma_{\bf k}g_{\bf k,
k'}\rho_{\bf k+k'}\theta_{\bf k}\theta_{\bf k'}
\\
H_{\rho^2} &\equiv& J\sum_{\bf k, k', q}(2-\gamma_{\bf
q})\nonumber\\
&&\times\left[\rho_{\bf -k-k'-q} \rho_{\bf q}-\rho_{\bf -k-q}\rho_{\bf
-k'+q}\right]\theta_{\bf k} \theta_{\bf k'}
\end{eqnarray}
where
\begin{equation}\label{Rho}
\rho_{\bf q} \equiv \frac{1}{N}\sum_{\bf r}e^{-i{\bf qr}}(c_{\bf
r}-c)\ ,
\end{equation}
\begin{equation}\label{def_of_g}
g_{\bf k, k'}\equiv{\textstyle\frac{\gamma_{\bf k+k'}-\gamma_{\bf
k}-\gamma_{\bf k'}}{\gamma_{\bf k}}}\ ,
\end{equation}
\begin{equation}
\gamma_{\bf k}\ =\ 2-\cos k_xa-\cos k_ya\ .
\end{equation}
The last relation is true for a square lattice with spacing $a$ and
the nearest neighbours interaction of strength $J$. It is
important to stress that the first term in the Hamiltonian can be
regarded as the SWA Hamiltonian of the model on a pure (undiluted)
lattice with a renormalised coupling. We can use it and write
thermodynamical averaging with respect to the Gibbs distribution
as

\begin{equation}
\left<...\right> = \left<...\ e^{-\beta(H_{\rho} +
H_{\rho^2})}\right>_*\ \left<e^{-\beta(H_{\rho} +
H_{\rho^2})}\right>_*^{-1}\ ,
\end{equation}
where the notation $\left<...\right>_*$ is used for
thermodynamical averaging with the Hamiltonian of the pure system:

\begin{eqnarray}
&&\left<...\right>_* = \frac{Tr \left(...\ e^{-c^2\beta
H_{\reg}}\right)}{Tr\ e^{-c^2\beta H_{\reg}}}\ ,
\\\nonumber
&&Tr\ (...)\equiv\left(\prod_{{\bf k}\in
B/2}\int_{-\infty}^{+\infty}d\theta_{\bf
k}^c\int_{-\infty}^{+\infty}d\theta_{\bf k}^s\right)(...)\ ,
\end{eqnarray}
where $\theta_{\bf r}^c\equiv \Re\theta_{\bf r}$, $\theta_{\bf
r}^s\equiv \Im\theta_{\bf r}$, and in order to keep the same
number $N$ of variables the product has to be taken over
a half of the 1st Brillouin zone which we have denoted as $B/2$ .
It was possible to extend the integration region to $(-\infty,+\infty)$
because of the Gaussian form of the Boltzman factor that stands in the
integrals.

\subsection{The expansion in $\{\rho_{\bf q}\}$}
\label{subsec2}

Let us note, that the transformation (\ref{Eq-HamDilute}) of the Hamiltonian
(\ref{EqH_dil}) is exact, although it looks like a perturbation expansion in
$\rho$. In the forthcoming calculations in order
to perform configurational averaging we expand any thermodynamical
quantity of interest $\left<F(\{\rho_{\bf q}\})\right>$ in terms
of functional variables $\rho_{\bf k}$'s, Eq.(\ref{Rho}):

\begin{eqnarray}
&&\left<F(\{\rho_{\bf q}\})\right> = \left<F(\{0\})\right> +
\sum_{\bf k}f_1({\bf k}) \rho_{\bf k} \label{Expns}
\\\nonumber
&& + \sum_{\bf k, k'}f_2({\bf k, k'}) \rho_{\bf k}\rho_{\bf k'} +
\sum_{\bf k, k', k''}f_3({\bf k, k', k''}) \rho_{\bf k}\rho_{\bf
k'}\rho_{\bf k''}+ \ ...
\end{eqnarray}

Since $\rho_{\bf k}$'s characterize the deviation from the completely
homogeneous disorder in the Hamiltonian they can be considered as
parameters of perturbation. Note that a power of $\rho$ corresponds
to the number of sums over ${\bf k}$ in (\ref{Expns}). A classification of the
perturbation theory series with respect to the number of sums over
${\bf k}$ corresponds to the expansion in the ratio of the volume
of effective interaction to the elementary cell volume
\cite{Vaks67}. Taking this ratio to be small means that it is
valid for the short-range interacting systems, which holds for our
problem. As far as we don't make any assumption about weakness of
disorder, we may expect that accordance of results of this expansion
with the MC simulations should not be very sensitive to the value of dilution
$(1-c)$ (but of
course still far enough from the percolation threshold where the
whole approach fails).

In the calculations presented below we limit ourselves to the
third order term in the expansion. Then it is not difficult to
perform averaging over configurations of disorder using the
equalities:

\begin{eqnarray}\nonumber
&&\overline{\rho_{\bf q}}\ =\ 0\ ,
\\\label{RhoAvrg}
&&\overline{\rho_{\bf q}\rho_{\bf q'}}\ =\ c(1-c)\frac{1}{N}
\delta_{{\bf q+q'},0}\ ,\label{Avrg}
\\\nonumber
&&\overline{\rho_{\bf q}\rho_{\bf q'}\rho_{\bf q''}}\ =\
c(1-3c+2c^2)\frac{1}{N^2}\delta_{{\bf q+q'+q''},0}
\end{eqnarray}
which can be obtained easily from (\ref{Eq-ConfAvg}).

\subsection{The asymptotic behaviour of the spin-spin correlation
function} \label{subsec3}

The spin-spin correlation function of the diluted 2D $XY$ model,

\begin{equation}
G_2(R) =\overline{\left<c_{\bf r}c_{\bf r+R}\cos(\theta_{\bf
r+R}-\theta_{\bf r})\right>}\ ,
\end{equation}
can be written in the Fourier variables (\ref{ThetaFour}) as
\begin{displaymath}
\overline{\left<c_{\bf r}c_{\bf
r+R}\cos\frac{1}{\sqrt{N}}\sum_{{\bf k}}\left( \eta^c_{\bf
k}\theta^c_{\bf k}+\eta^s_{\bf k}\theta^s_{\bf k}\right)\right>}
\end{displaymath}
with
\begin{eqnarray}\nonumber
&&\eta^c_{\bf k}\ =\ \cos{\bf kr}-\cos{\bf k(r+R)}\ ,
\\
&&\eta^s_{\bf k}\ =\ -(\sin{\bf kr}-\sin{\bf k(r+R)})\ .
\end{eqnarray}
Writing the expansion (\ref{Expns}) and applying the equalities
(\ref{Avrg}) we arrive (see appendix A) at the next expression:

\begin{eqnarray}\nonumber
G_2(R) &=& c^2 \left<\cos(\theta_{\bf r+R}-\theta_{\bf
r})\right>_*
\\\nonumber
&\times&\Bigg[\ 1\ - \frac{1-c}{c^3}\ \frac{1}{\beta J}
\frac{1}{N^2}\sum_{{\bf k},{\bf k'}}g_{\bf k, k'}g_{\bf k',
k}{\textstyle\frac{\sin^2\frac{\bf kR}{2}}{\gamma_{\bf k}}}
\\\nonumber
&&+\ \frac{1-3c+2c^2}{c^4}\ \frac{1}{\beta J}
\Bigg(\frac{2}{N}\sum_{{\bf k}}{\textstyle\frac{\sin^2\frac{\bf
kR}{2}}{\gamma_{\bf k}}}
\\\label{G}
&& - \frac{1}{N^3}\sum_{{\bf k},{\bf k'},{\bf k''}}g_{\bf -k,
k'}g_{\bf k', k''}g_{\bf k'', k} {\textstyle\frac{\sin^2\frac{\bf
kR}{2}}{\gamma_{\bf k}}} \Bigg) \Bigg]\ .\phantom{****}
\end{eqnarray}

Since Eq.(\ref{G}) is already configurationally averaged it does not
contain the $\rho$'s anymore, so the correspondence with the orders of
the expansion (\ref{Expns}) is not obvious.
Let us explain the origin of each term. The
unity corresponds to the zeroth-order in $\rho$, the first-order
term is identically vanishing as follows from (\ref{RhoAvrg}), and
the second- and third-order
terms in $\rho$ can be distinguished by their coefficients that
are clear from (\ref{RhoAvrg}). The second-order term contains two sums
according to the expansion (\ref{Expns}). At the same time the third-order
term contains, except the triple sum, a sum over one ${\bf k}$:
the summation over two remaining ${\bf k}$'s was possible to carry out
explicitly in this particular case.

For our purpose it is enough to get the leading asymptotics of the sums
that stand in the expression (\ref{G})
when $N\rightarrow\infty$ and
$R\rightarrow\infty$ (see Appendix B):

\begin{eqnarray}\nonumber
&&\frac{1}{N}\sum_{\bf k}{\textstyle\frac{\sin^2\frac{\bf kR}{2}}
{\gamma_{\bf k}}}\ \approx\ {\rm const}\ +\
\frac{1}{2\pi}\ln{\textstyle\frac{R}{a}}\ ,
\\\nonumber
&&\frac{1}{N^2}\sum_{\bf k, k'}g_{\bf k, k'}g_{\bf k', k}
{\textstyle\frac{\sin^2\frac{\bf kR}{2}}{\gamma_{\bf k}}}\
\approx\ {\rm const}'\ +\
\frac{0.73}{2\pi}\ln{\textstyle\frac{R}{a}}\ ,
\\\nonumber
&&\frac{1}{N^3} \sum_{{\bf k},{\bf k'},{\bf k''}} g_{\bf -k,
k'}g_{\bf k', k''}g_{\bf k'', k} {\textstyle\frac{\sin^2\frac{\bf
kR}{2}}{\gamma_{\bf k}}} \approx {\rm const}'' - \frac
{0.27}{2\pi}\ln{\textstyle\frac{R}{a}}.
\end{eqnarray}
Inserting these expressions in (\ref{G}) it is possible to write
the pair correlation function for small enough temperatures in the
power law form:
\begin{equation}
G_2(R) \approx c^2 (R/a)^{-\eta^{\dil}}\ .
\end{equation}
Reminding the spin-spin correlation function exponent of the pure
system, $\eta^{\reg}$, given in the SWA by Eq.(\ref{EtaReg}), we write
\begin{equation}
\eta^{\dil} = \eta^{\reg}\left( \frac {1}{c^2}+0.73\frac
{1-c}{c^3}-2.27\frac{(1-3c+2c^2)}{c^4} \right)\ .
\label{Eq-etabetter}
\end{equation}
In fact, as it can be seen from Appendix B, this result is true
not only in the thermodynamic limit but also for a system of finite
large enough size $N$.
The first term in the brackets, $1/c^2$, corresponds to the
zeroth order of the $\rho$-expansion, the first-order
term is identically vanishing as was already noted before, the
second and third terms in the brackets correspond to the second-
and third-order terms of the $\rho$-expansion respectively. In the next
subsection we evaluate formula (\ref{Eq-etabetter}) and compare this
result with the MC experiments.

\subsection{Comparison with the Monte Carlo results for the exponent of the
spin-spin correlation function}\label{subsec4}

In order to check Eq.(\ref{Eq-etabetter}), we have performed
simulations of 2D XY-spins using Wolff cluster Monte Carlo
algorithm~\cite{Wolff89}.
We only mention here the main features of the
simulations  used in order to obtain the exponent $\eta^{\dil}$.
We discard typically $10^5$ sweeps for thermalization, and the
measurements are performed with typically $10^5$ production sweeps.
Averages over disorder are performed using typically $10^3$ samples.
There is no need of a better statistics, since we are far from the BKT point
(in the vicinity of the deconfining transition, the presence of many
topological defects is an obstacle to thermalization as it can be shown
empirically by the analysis of autocorrelation time (see e.g.
Ref.~\cite{FarinasEtAl05})).
The boundary conditions are chosen periodic and the
critical exponent $\eta(T)$ of the correlation function is
measured indirectly through the finite-size scaling behaviour of
the magnetisation \be\label{FinSizeSc} M_T(L)\sim L^{-x_\sigma(T)}\ ,\quad
x_\sigma(T)=\frac 12\eta(T), \ee where the last scaling relation holds in
two dimensions ($L=\sqrt{N}$ is the linear size of the lattice).

\begin{figure}[ht]
\hspace{0cm}
\resizebox{0.95\columnwidth}{!}{%
\includegraphics{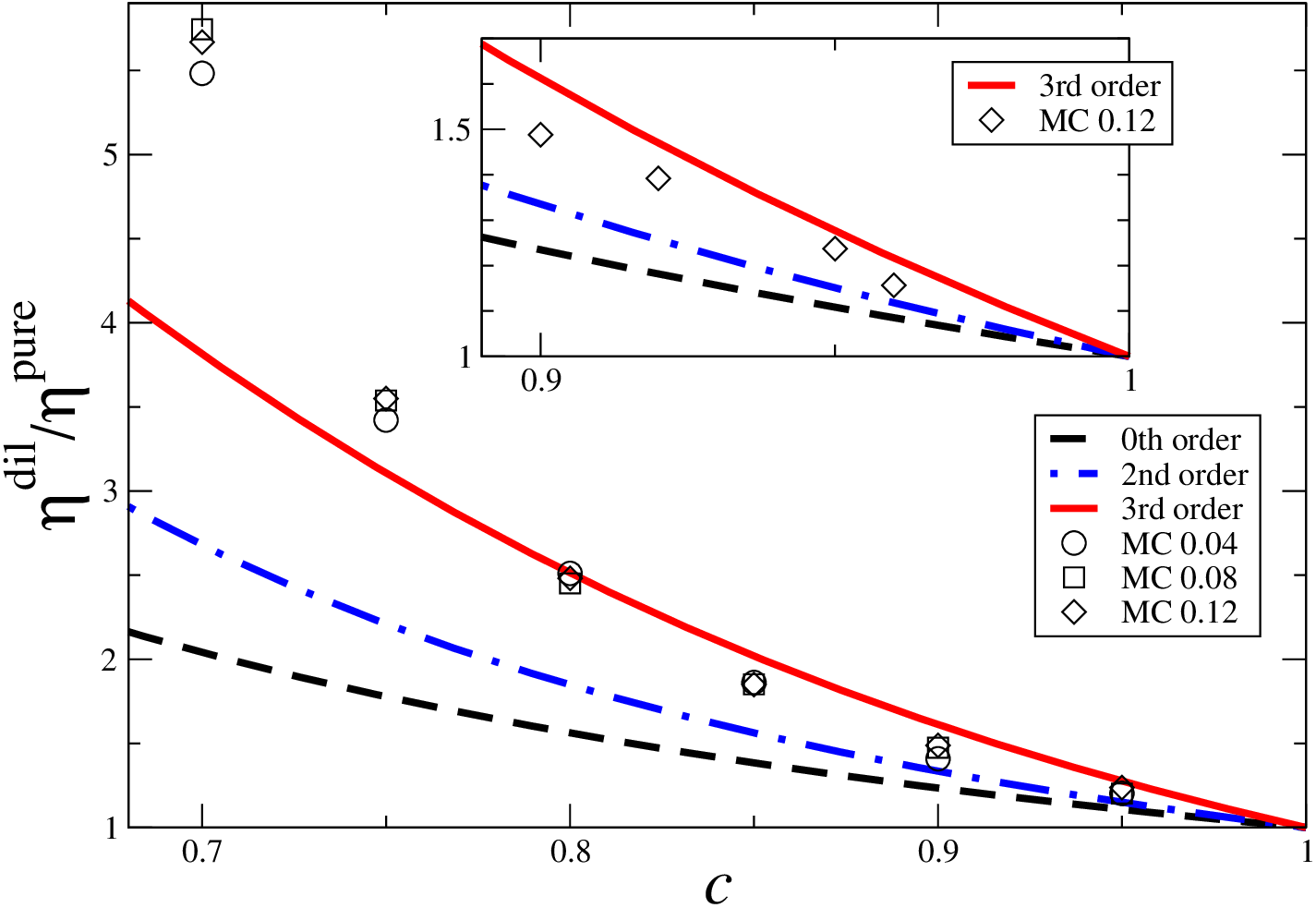}}
\caption{The comparison between the MC data and the analytic results
for the ratio $\eta^{\dil}/\eta^{\reg}$ as a function of
concentration of occupied sites $c$ obtained in different
orders of the $\rho$-expansion.}
\label{Fig1}       
\end{figure}

In Fig.~\ref{Fig1}, we compare the ratio $\eta^{\dil}/\eta^{\reg}$
evaluated analytically by keeping terms from the 1st to 3rd order in $\rho$-expansions
with the MC data. One can see that up to the third order the analytic
curves approach step by step the MC data: the 0th order seems to be a
rough approximation, the 2nd order curve lies closer to the MC data, but
still much below, and the third order curve seems to fit the MC results
better, although it doesn't give perfect accordance.
Of course this fact does not allow
to conclude in favour of a similar
agreement for  higher orders, and it is
possible that the expansion will not show any convergence at all.
On the other hand, having more perturbation theory contributions at hand
one can attempt to apply a resummation technique to improve its convergence,
similarly as it is commonly done analyzing field-theoretical expansions
~\cite{Zinn-Justin}.

Anyway, comparing outcomes of our analytical and MC treatments presented
in Fig.\ref{Fig1} one arrives at the conclusion about good agreement within
concentrations from $c=0.75$ to $c=1$ at least up to the third order of the
perturbation expansion.

Let us proceed further investigating magnetization and its distribution in
a finite-size system.

\section{The magnetisation probability distribution}

In this section we obtain and discuss the probability distribution
function of magnetisation obtained in Monte Carlo simulations of a
two-dimensional $XY$-spin system, performed for different sizes of
the lattice, temperatures and concentrations of impurities. We
support the MC analysis with an analytic treatment of the
magnetisation probability distribution in a model of finite size
using the same approach as for the spin-spin correlation function.

\subsection{The probability distribution functions}

In subsection \ref{subsec4} we obtained the spin-spin correlation function decay
exponent by use of the finite size scaling relation (\ref{FinSizeSc})
for the magnetisation measured in MC simulations. At the same time the
probability distribution function (PDF) of magnetisation itself deserves much attention
since it appears to be of non-trivial form. As we mentioned in Introduction
it is known that it is non-Gaussian in the pure 2D $XY$ model. In the case of
structural disorder we can expect also dependence on concentration of dilution.

We define instantaneous magnetisation as the scalar value of the
total sum of the spins divided by the number of sites,

\begin{equation}\label{magnetis}
m = \frac{1}{N}\left|\sum_{\bf r}c_{\bf r}{\bf S}_{\bf r}\right|\ ,
\end{equation}
measured in a given state from the thermodynamical ensemble of states of
the system with a fixed  configuration of structural disorder. The probability
to find the system in a state with magnetisation
$m$, $P_{\conf}(m)$, considered as a function of $m$ is called the probability distribution
function (PDF) of magnetisation or just the magnetisation probability distribution.

The thermodynamical mean value
of magnetisation defined through the usual procedure
of thermodynamical averaging with the Hamiltonian of the
system $H$:

\begin{equation}\label{magnetis3}
\left<m\right> = \frac{{\rm Tr}\left(m e^{-\beta H}\right)}
{{\rm Tr}\ e^{-\beta H}}\ ,
\end{equation}
can be written then in terms of the PDF as

\begin{equation}\label{magnetis4}
\left<m\right> = \int_{0}^{1}m P_{\conf}(m)dm\ .
\end{equation}
We define also the $p$th moment of magnetisation as

\begin{equation}
M_p \equiv \left<m^p\right> = \int_{0}^{1}m^pP_{\conf}(m)dm\ .
\end{equation}

In this place it is important to emphasize that the magnetization
$m$, which stands in the integral in (\ref{magnetis4}), differs from
that defined by Eq.(\ref{magnetis}), since in (\ref{magnetis4}) $m$
just plays the role of the integration variable which doesn't depend
on the microscopic state of the system. The PDF, $P_{\conf}(m)$, is
already a thermodynamic characteristic depending only on a
macroscopic state of the system. It can be seen from the well known
property of probability distribution
functions that a PDF is defined uniquely by its
moments~\cite{Gnedenko}:

\begin{equation}\label{PDFinMoments}
P_{\conf}(m) =
\int_{-\infty}^{\infty}\frac{dx}{2\pi}e^{imx}\sum_{p=0}^{\infty}
\frac{(-ix)^p}{p!}M_p\ .
\end{equation}
The moments are thermodynamically averaged quantities, so it follows
that the PDF of magnetization is a thermodynamic quantity too and
depends on $m$ only as on a parameter, however to find an analytic
expression for the PDF is not a trivial task even for the pure
model.

Thus the mean magnetisations defined by Eq.(\ref{magnetis3}) and Eq.(\ref{magnetis4})
being the same are obtained by different procedures.

Since we investigate observable quantities here we should look at the configurationally
averaged values of magnetisation and its moments. In terms of the PDF it means that
$P_{\conf}(m)$ must be averaged over the whole range of possible configurations of impurities
with a fixed concentration. Then the mean magnetisation and its moments can be written as:

\begin{equation}\label{magnetis2}
\overline{\left<m\right>} = \int_{0}^{1}m P(m)dm
\end{equation}
and

\begin{equation}
\overline{M_p} = \int_{0}^{1}m^pP(m)dm\ ,
\end{equation}
where $P(m)=\overline{P_{\conf}(m)}$ is the configurationally
averaged PDF.

\begin{figure}[ht]
\hspace{0cm} \resizebox{0.95\columnwidth}{!}{
\includegraphics{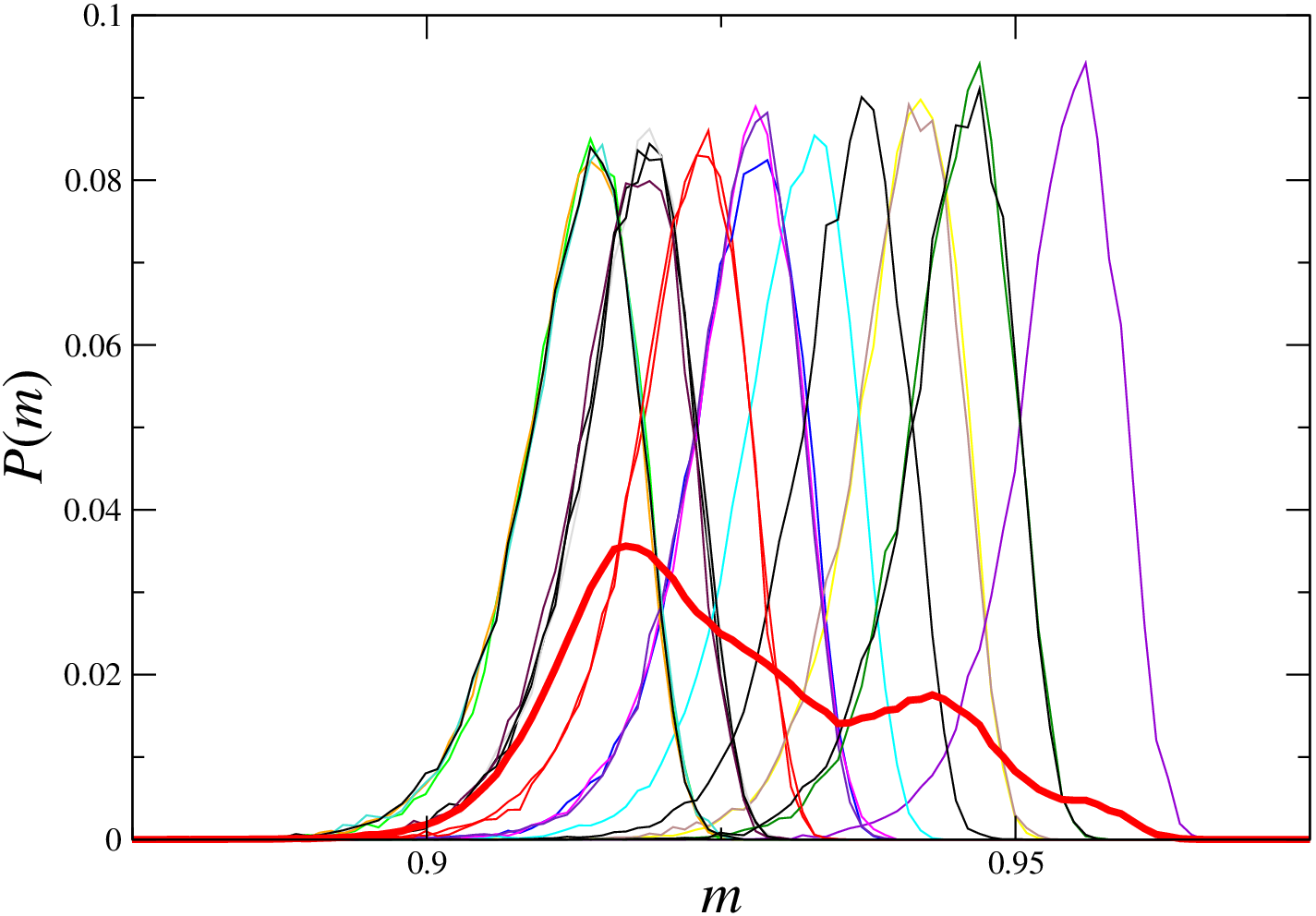}}
\caption{The probability distributions of magnetisation
    for twenty different realizations for a system of size $L=16$
    at dilution $c=0.95$ at a temperature  $k_BT/J=0.1$. The thick line
    is the average probability distribution, still very
    bumpy with so few configurations.}\label{Fig2}
\end{figure}

The PDF of magnetization  is very suitable
for further analysis of results obtained in Monte Carlo simulations.
In Fig.~\ref{Fig2} we
illustrate the procedure of configurational averaging of MC data. From
different curves of $P_{\conf}(m)$ obtained for different realizations of dilution
we draw one averaged curve which is $P(m)$.

\begin{figure}[ht]
\hspace{0cm} \resizebox{0.95\columnwidth}{!}{
\includegraphics{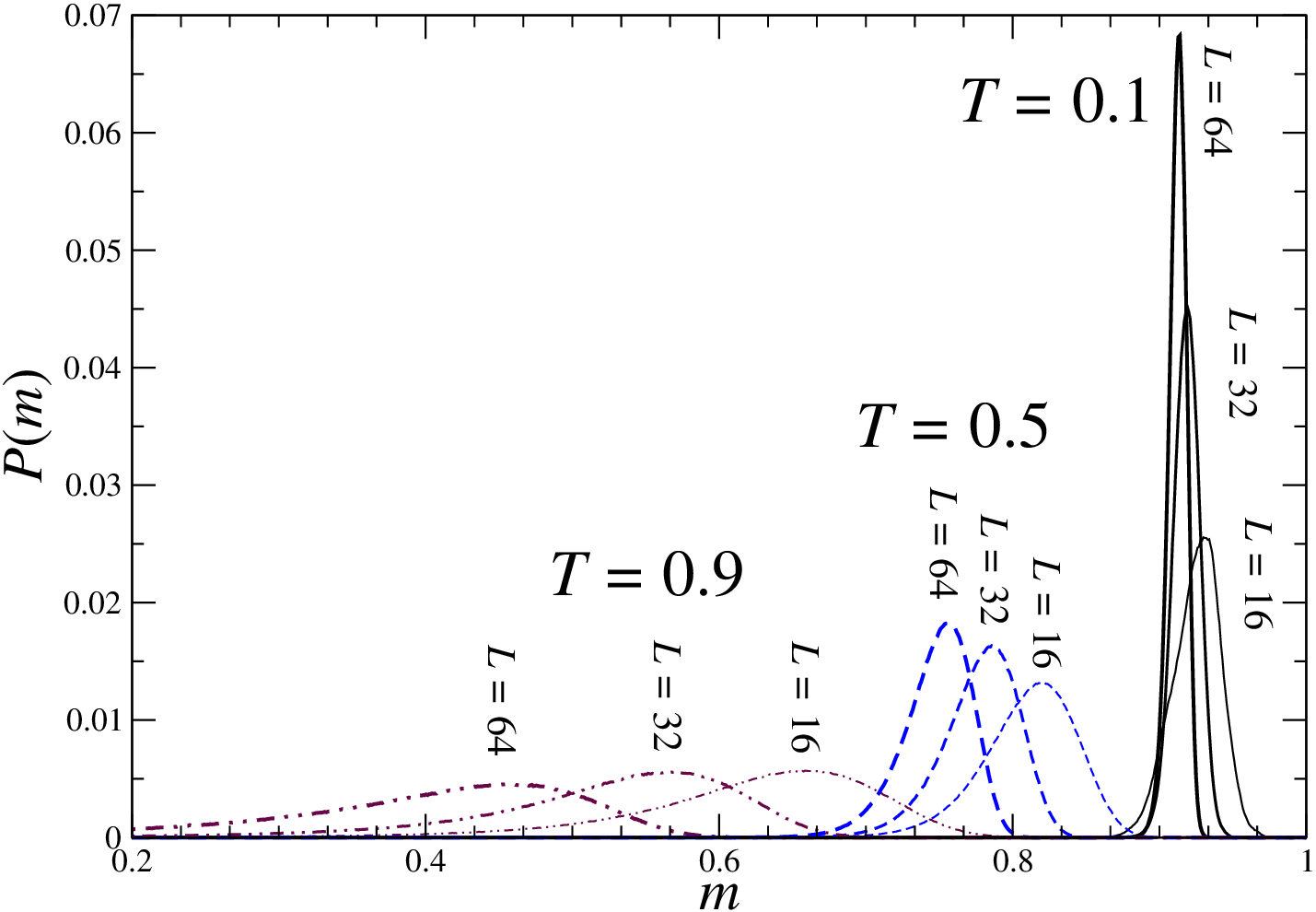}}
\caption{The average distributions over $10^3$ samples at
temperatures
    $k_BT/J=0.1$,  0.5 and 0.9 for systems of increasing
    sizes $L=16$, 32, 64. }\label{Fig3}
\end{figure}

In the pure 2D $XY$ model two of the main features of the PDF are
that the form of the distribution at fixed temperature is universal,
i. e. it does not depend on the size of the system, and it is non-Gaussian.
These two statements have been derived analytically and verified by
MC simulations~\cite{Archambault97,BanksBramwell05,BramwellEtAl01}.

Fig.~\ref{Fig3} illustrates the MC results for size dependence of the PDF of
magnetisation in a diluted 2D $XY$-spin system with concentration of spins
$c=0.95$ at three different temperatures. We see that at fixed
temperature the mean
magnetisation becomes smaller for bigger lattices as it should be
according to Eq.(\ref{MagnDecay}) in the case of a pure system.
From the curves for low temperatures
it is clear that the form of the distributions is non-Gaussian
like in the pure model.
What is more interesting is that the form of the
distributions is noticeably different for
different sizes. It seems to be in contradiction
with results for the pure model
~\cite{Archambault97,BanksBramwell05,BramwellEtAl01}.

A suitable parameter that can characterize the form
of a PDF is the variance: $\sigma=\sqrt{\overline{M_2}-\overline{M_1}^2}$.
It has been prooved that in the pure 2D $XY$ model
it is independent of system size. Here we see in Fig.\ref{Fig3}
a different qualitative behaviour,
the variance, which is proportional to width and
flatness of the distribution, grows as the size of the lattice
decreases.

Since we simulated a diluted system, this
must be the result of non-magnetic
impurities influence. It calls for an analytic
explanation of this dependence in Subsection 3.3.

\subsection{The ring functions}

\begin{figure}[!htb]

\hspace{1cm} \resizebox{0.75\columnwidth}{!}{
\includegraphics{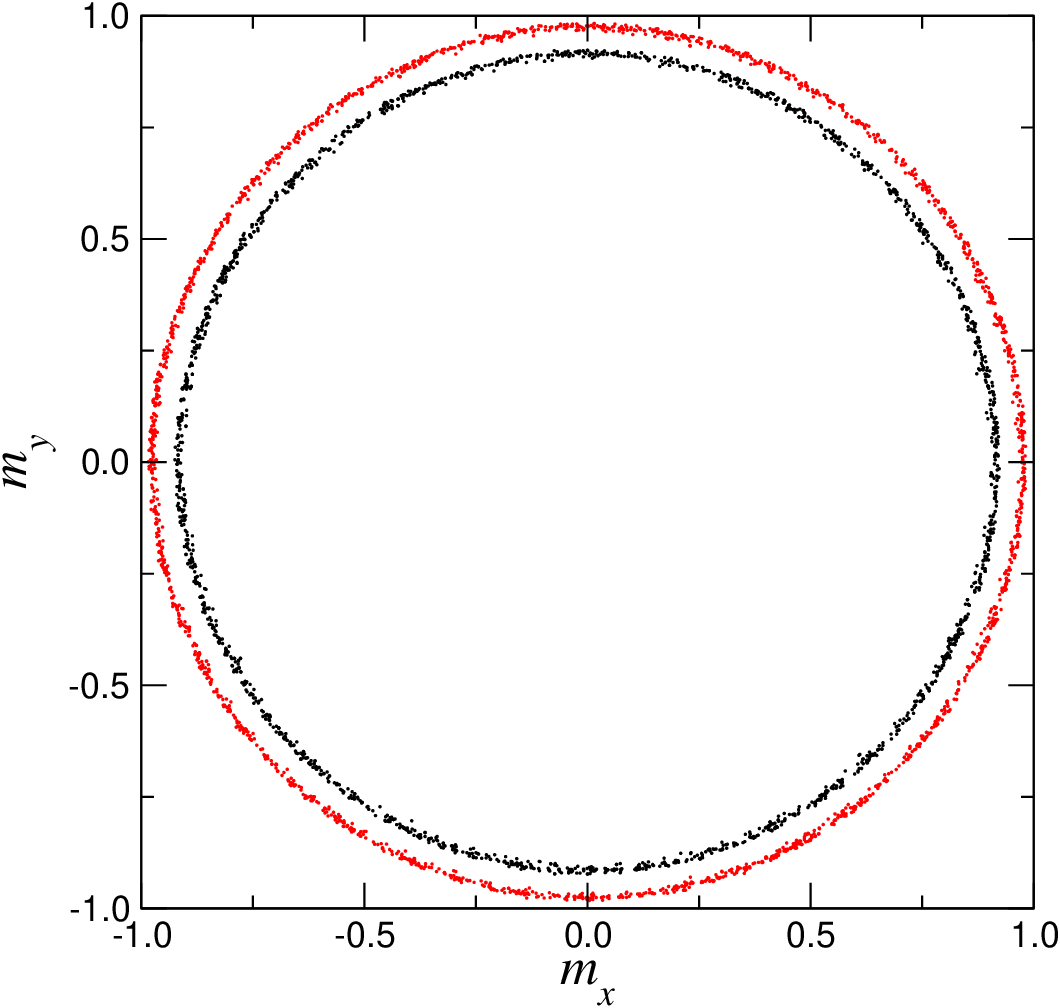}}

\hspace{1cm} \resizebox{0.75\columnwidth}{!}{
\includegraphics{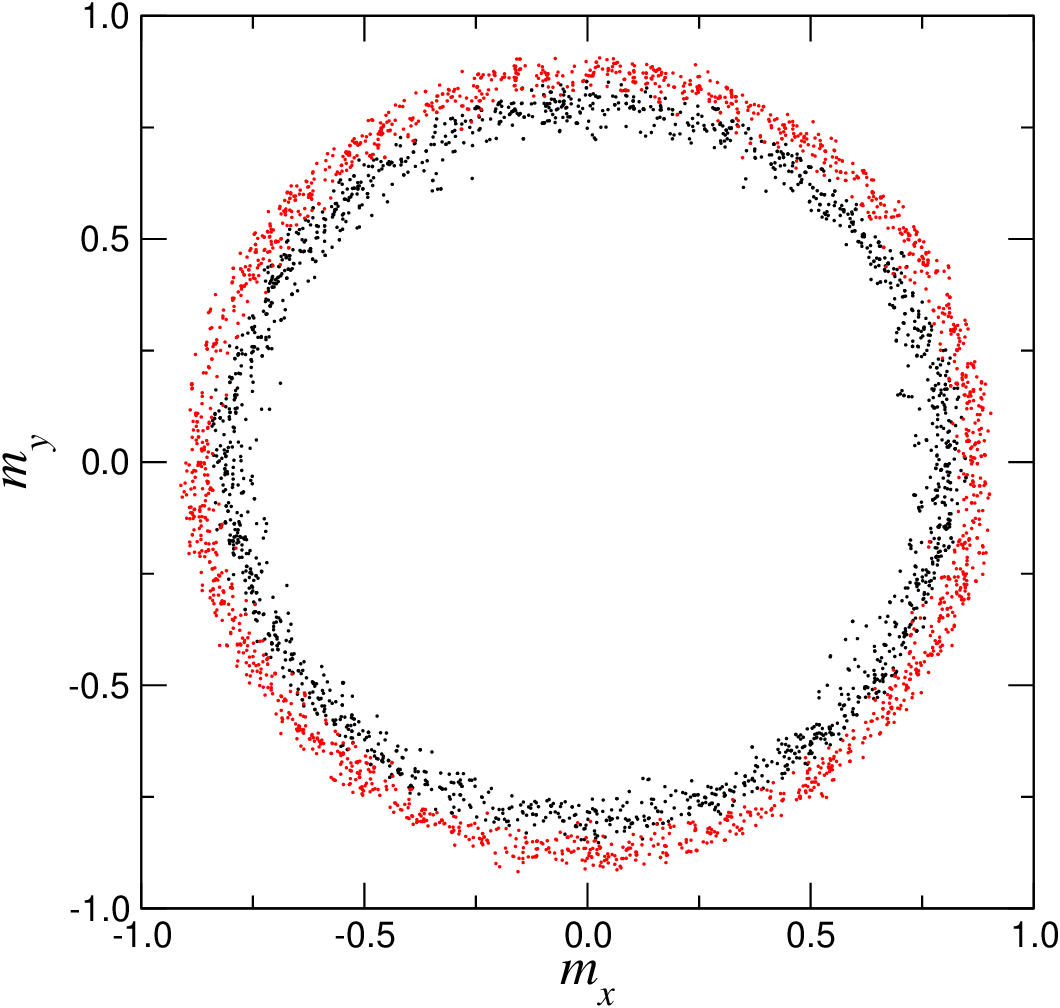}}

\hspace{1cm} \resizebox{0.75\columnwidth}{!}{
\includegraphics{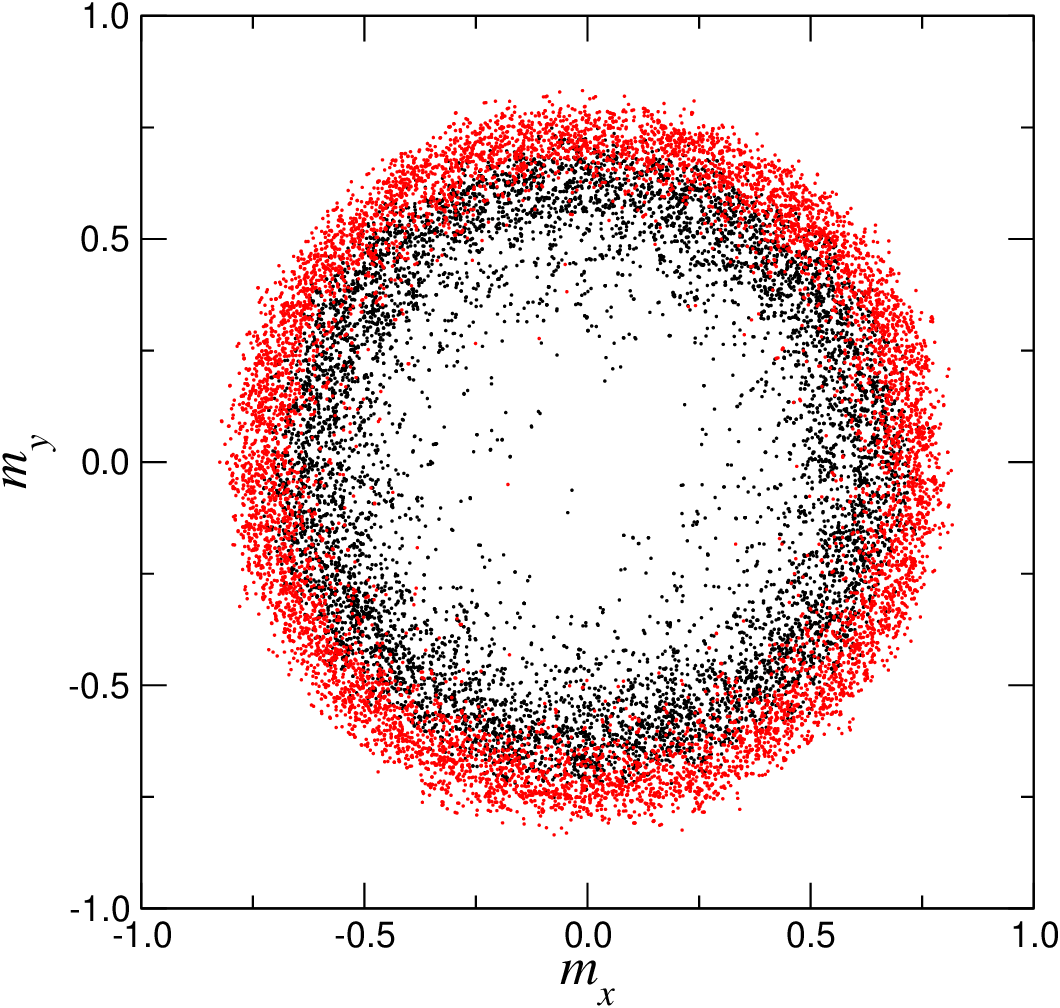}}

    \vspace{2mm}
  \caption{Ring functions for a system of size $L=16$ at dilution $c=0.95$
    for temperatures (from top to bottom) $k_BT/J=0.1$,
0.5, and 0.9.
    The outer ring function (color on-line) represents the pure system.}
  \label{Fig4}
\end{figure}

Another way to display the magnetisation probability distribution
observed in MC simulations is to draw a ring function which is defined in the next way.
A ring function is obtained when one plots the successive
values of the magnetisation (for each Monte Carlo step) in the
plane $(m_x,m_y)$ where $m_x$ and $m_y$ are the two components of
the magnetisation (see Fig.\ref{Fig4}). In fact it contains the same
information as a PDF, the difference is that in a ring function
we can see the distribution of the {\em vector} of magnetisation,
its form however is
symmetric around the point $m=0$, i. e. there is no
preferable direction (this is the signature of rotational invariance of
the QLRO phase),
and the probability, $P(m_x,m_y)$, is mirrored in density of the points.

Since the algorithm used here is a cluster algorithm specially
dedicated to this type of spin systems, the successive spots are
essentially uncorrelated. This is a very different situation with
a Metropolis algorithm where the successive spots would be
correlated (see Ref. \cite{Archambault97}).

We are interested in the temperature dependence of a ring function
of magnetisation in a diluted 2D $XY$-spin system with fixed size
and concentration of impurities. For this purpose ring functions
for a system of size $L=16$ at dilution $c=0.95$ for three
different temperatures, $k_BT/J=0.1$, 0.5 and 0.9, are shown in
Fig.~\ref{Fig4}. The outer ring functions (color on-line)
represent the pure system of the same size and at the same
temperatures. The first feature that one can notice is that the
radius of the rings of higher density of the ``diluted" ring
functions is smaller for all temperatures than that of the
corresponding ``pure'' ring functions. This is due to the fact
that we consider the magnetisation per site taking all sites into
account and in a system with impurities there are missing spins.

In which concerns the temperature dependence of the ring function,
it is visible that the mean magnetisation tends to zero as the
temperature increases in both pure and diluted system. When for
the pure 2D $XY$ model this behaviour follows from
Eq.(\ref{MagnDecay}) (since $\eta^{\reg}\sim k_BT$), it must be
verified analytically for the case of a model with site-dilution,
what is done in Subsection 3.3. Another feature of the
temperature-behaviour which can be noticed in Fig.\ref{Fig4} is
that the high-density region is wider for higher temperatures and
approaches to a delta-function as the temperature goes to zero,
this feature is well known from MC and analytic investigations of
the pure 2D $XY$ model~\cite{Archambault97,BramwellEtAl01}.

Except the difference in position of the density peaks of the
ring functions caused by the difference in the total numbers of spins,
there are no qualitative differences in the form of the ``diluted" and
``pure" ring functions obtained at the same temperature for the value of
concentration of magnetic sites $c=0.95$. Their widths are approximately
the same and both distributions show clear non-Gaussian character for
higher temperatures
which lies in the visible fact that more points are situated in the inner
region of the ring functions than out of the rings of the highest density.

\begin{figure}[ht]
\hspace{1cm}
\resizebox{0.75\columnwidth}{!}{
\includegraphics{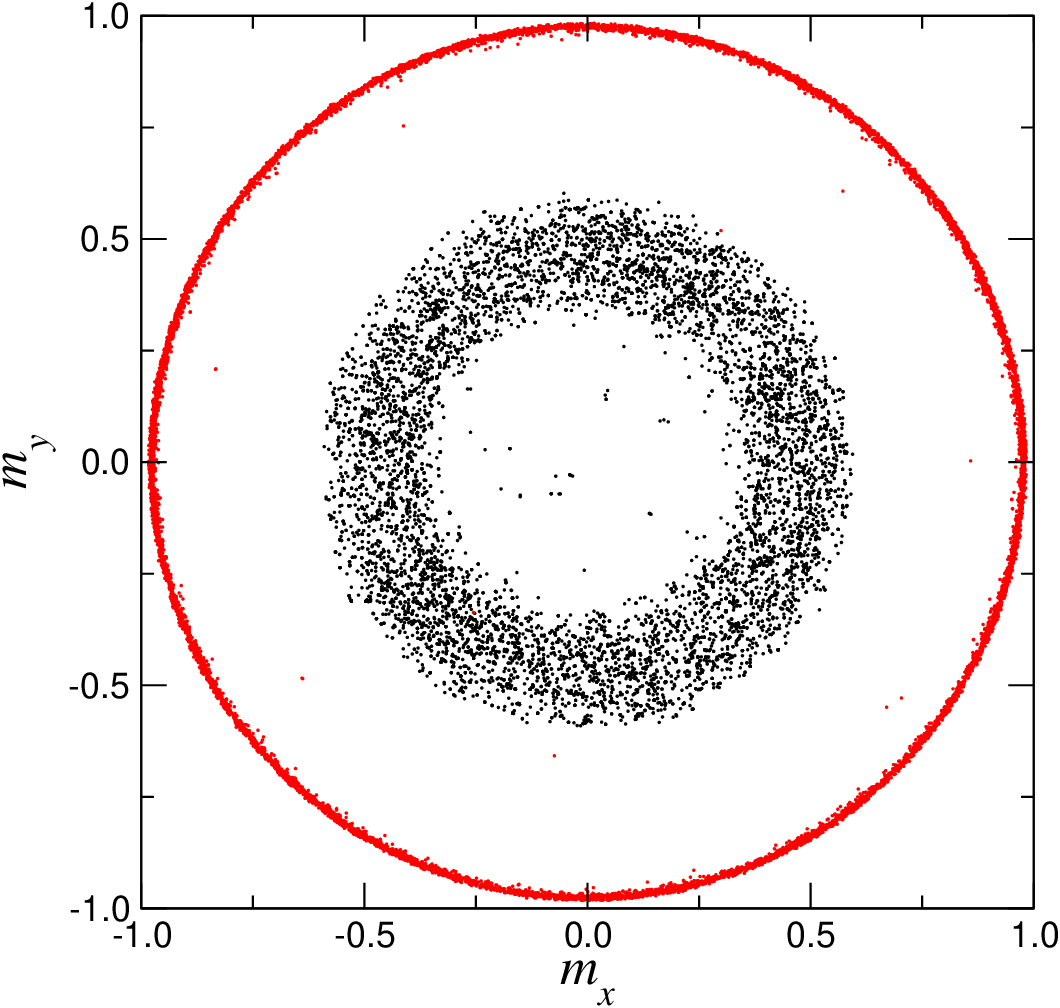}}
\caption{A ring function for a system of size $L=16$ at
dilution  $c=0.70$ for temperature $k_BT/J=0.1$. The outer ring
function represents the pure system.}\label{Fig5}
\end{figure}

The discussion above concerns  relatively weak dilution with
$c=0.95$. The situation appears quite different when the dilution
becomes stronger.
In Fig.~\ref{Fig5}
we show a ring function for a system of the same size as
in the previous figure at
temperature $k_BT/J=0.1$ but with much stronger dilution, $c=0.70$.
One can see now that in a system with sufficient number of non-magnetic
impurities the high-density region of the
ring function is much wider than in a pure system at the
same temperature. We can conclude thus that the variance that
characterizes the width of the distribution must be dependent on
concentration of dilution in a 2D $XY$ model with structural
disorder.
This non-trivial observation calls for an analytic
support which is the point of the next subsection.

\subsection{Analytic calculation of magnetisation and its moments}

In this subsection we present an analytic analysis of the PDF of magnetisation
in the 2D $XY$ model in the case of structural disorder. We use here the same scheme
that was presented in Section \ref{sec1} to obtain the moments of magnetisation
which define the magnetisation probability distribution through Eq.(\ref{PDFinMoments}).

It is convenient for analytic calculation to rewrite the
instantaneous magnetization (\ref{magnetis}) as

\begin{equation}
m=\frac{1}{N}\sum_{\bf r} c_{\bf r}\cos(\theta_{\bf r}-\overline{\theta})\ ,
\end{equation}
where $\overline{\theta}\equiv\frac{1}{N}\sum_{\bf
r}\theta_{\bf r}$ is the algebraic mean of the angle variables.

Since we are interested in observable quantities we must consider
configurationally averaged moments of magnetization:

\begin{displaymath}
\overline{M_p}\equiv\overline{\left<m^p\right>}
=\frac{1}{N^{p}}\sum_{\bf r_1,...r_p}\overline{\left<c_{\bf
r_1}\cdot\cdot\cdot c_{\bf r_{p}}\cos\psi_{\bf
r_1}\cdot\cdot\cdot\cos\psi_{\bf r_p}\right>}\ ,
\end{displaymath}
where we denoted $\psi_{\bf r}=\theta_{\bf r}-\overline{\theta}$.
Note here that the 1st moment of magnetisation,
\begin{displaymath}
\overline{M_1} = \frac{1}{N}\sum_{\bf r}
\overline{\left<c_{\bf r}\cos\psi_{\bf r}\right>}=
\overline{\left<c_{0}\cos\psi_{0}\right>}\ ,
\end{displaymath}
is just the mean magnetisation $\overline{\left<m\right>}$ by
definition.

Passing from the product of cosines to a sum we write
the $(n+1)$th moment as

\begin{eqnarray}\label{moment}
&&\overline{M_{n+1}}
\\\nonumber
&&=\frac{1}{2^{n}N^{n}}\sum_{\bf r_1,... r_n}\sum_{\alpha_i=\pm 1}
\overline{\left<c_0 c_{\bf r_1}\cdot\cdot\cdot 
\cos{\textstyle\left(\psi_0+\sum_{i=1}^n\alpha_i\psi_{\bf
r_i}\right)}\right>}\ .
\end{eqnarray}
The expression that stands in the sums in
(\ref{moment}) can be written in Fourier variables as:

\begin{eqnarray}\nonumber
&&\overline{\left<c_0 c_{\bf r_1}\cdot\cdot\cdot c_{\bf
r_n}\cos(\psi_0+\alpha_1\psi_{\bf
r_1}\cdot\cdot\cdot+\alpha_n\psi_{\bf r_n})\right>}
\\\nonumber
&&=\ \overline{\left<c_0c_{\bf r_1}\cdot\cdot\cdot c_{\bf
r_n}\cos{\textstyle\frac{1}{\sqrt{N}}\sum_{{\bf k}}\left(
\eta^c_{\bf k}\theta^c_{\bf k}+\eta^s_{\bf k}\theta^s_{\bf
k}\right)}\right>}
\end{eqnarray}
with
\begin{eqnarray}\nonumber
&&\eta^c_{\bf k}\ =\ 1+\alpha_1\cos{\bf
kr_1}+\cdot\cdot\cdot+\alpha_n\cos{\bf kr_n}\ ,
\\
&&\eta^s_{\bf k}\ =\ -\left(\alpha_1\sin{\bf
kr_1}+\cdot\cdot\cdot+\alpha_n\sin{\bf kr_n}\right)\ .
\end{eqnarray}

Substituting in (\ref{moment}) the result for the above expression
obtained in Appendix A in
the third order approximation in the $\rho$-expansion, Eq.(\ref{appAresult}),
we get
\begin{eqnarray}\nonumber
&&\overline{M_{n+1}} = \frac{c^{n+1}}{2^{n}N^{n}}\sum_{\bf r_1,...
r_n}\sum_{\alpha_i=\pm 1}
\left<\cos{\textstyle(\psi_0+\sum_{i=1}^n\alpha_i\psi_{\bf
r_i})}\right>_*
\\\nonumber
&&\times\ \Bigg[\ \ 1 - \frac{1}{\beta J}\Bigg(
\frac{1-c}{4c^3}\frac{1}{N^2}\sum_{{\bf k},{\bf
k'}}{\textstyle\frac{g_{\bf k, k'}g_{\bf k', k}}{\gamma_{\bf
k}}}+\frac{1-3c+2c^2}{c^4}
\nonumber\\ \label{moment1}
&&\times\Bigg(\frac{1}{2N}\sum_{{\bf
k}}{\textstyle\frac{1}{\gamma_{\bf k}}}-\frac{1}{4N^3}\sum_{{\bf
k},{\bf k'},{\bf k''}}g_{\bf k, k'}g_{\bf k', k''}g_{\bf k'',
k}{\textstyle\frac{1}{\gamma_{\bf k}}}\Bigg) \Bigg)
\nonumber\\
&&\textstyle
\qquad\times\left(n+1+2\sum_{i<j}\alpha_i\alpha_j\cos{\bf
k(r_i-r_j)}\right)\ \Bigg] \label{Eq-Correl}
\end{eqnarray}
with $g_{\bf k, k'}$ given by Eq.(\ref{def_of_g}).
Then, using the result for a quantity of the type
$\left<\cos{\textstyle\frac{1}{\sqrt{N}}\sum_{{\bf k}}\left(
\eta^c_{\bf k}\theta^c_{\bf k}+\eta^s_{\bf k}\theta^s_{\bf
k}\right)}\right>_*$ from Appendix A, Eq.(\ref{CosMean}),
we write in the low-temperature limit
\begin{eqnarray}
&&\left<\cos(\psi_0+\alpha_1\psi_{\bf
r_1}\cdot\cdot\cdot+\alpha_n\psi_{\bf r_n})\right>_*
\ \approx\ 1
\\\nonumber
&& - {\textstyle\frac{1}{4c^2\beta JN}}\sum_{{\bf k}\neq
0}{\textstyle\frac{1}{\gamma_{\bf
k}}\left(n+1+2\sum_{i<j}\alpha_i\alpha_j\cos{\bf
k(r_i-r_j)}\right)}\ .
\end{eqnarray}
Substituting this expression in (\ref{moment1}) we are able to sum
over all $\alpha_i$'s by help of the obvious equalities:

\begin{displaymath}
\sum_{\alpha_i=\pm 1}\alpha_i=0\ ,\quad\alpha_i^2=1\ .
\end{displaymath}
Then the $(n+1)$th moment of magnetisation reads as:
\begin{eqnarray}\label{moment3}
&&\overline{M_{n+1}} = c^{n+1}\Bigg[1 - \frac{n+1}{\beta
J}\Bigg(\frac{1}{4c^2}\frac{1}{N}\sum_{{\bf k}\neq
0}{\textstyle\frac{1}{\gamma_{\bf k}}}
\\\nonumber
&& -\frac{1-c}{4c^3}\frac{1}{N^2}\sum_{{\bf k},{\bf k'}}g_{\bf k,
k'}g_{\bf k', k}{\textstyle\frac{1}{\gamma_{\bf
k}}}+\frac{1-3c+2c^2}{2c^4}
\\\nonumber
&&\times\Bigg(\frac{1}{N}\sum_{{\bf
k}}{\textstyle\frac{1}{\gamma_{\bf k}}} -\frac{1}{2N^3}\sum_{{\bf
k},{\bf k'},{\bf k''}}g_{\bf -k, k'}g_{\bf k', k''}g_{\bf k'',
k}{\textstyle\frac{1}{\gamma_{\bf k}}} \Bigg)\Bigg) \Bigg].
\end{eqnarray}
In the limit $N\rightarrow\infty$ we find for the sums in (\ref{moment3})
(see Appendix B):
\begin{eqnarray}\nonumber
&&\frac{1}{N}\sum_{\bf k}{\textstyle\frac{1}{\gamma_{\bf k}}}
\approx {\rm const}+ \frac{1}{2\pi}\ln N,
\\\nonumber
&&\frac{1}{N^2}\sum_{\bf k}\sum_{\bf k'}g_{\bf k, k'}g_{\bf k',
k}{\textstyle\frac{1}{\gamma_{\bf k}}} \approx {\rm const}' +
\frac{0.73}{2\pi}\ln N,
\\\nonumber
&&\frac{1}{N^3} \sum_{{\bf k},{\bf k'},{\bf k''}}g_{\bf -k,
k'}g_{\bf k', k''}g_{\bf k'', k}{\textstyle\frac{1}{\gamma_{\bf k}}}
\approx {\rm const}'' - \frac{0.27}{2\pi}\ln N\ .\nonumber
\end{eqnarray}
For small enough temperatures it is now possible to write the $p$th
moment of magnetisation in the form:
\begin{equation}\label{moment4}
\overline{M_{p}} \approx c^{p}N^{-\frac{p}{4}\eta^{\dil}}
\end{equation}
with the exponent $\eta^{\dil}$ given by Eq.(\ref{Eq-etabetter}).
The equality (\ref{moment4}) can be rewritten as

\begin{equation}\label{moment5}
\overline{M_{p}} \approx \overline{M_{1}}^{p}\ .
\end{equation}
As far as all higher moments $\overline{M_{p}}$ can be trivially
expressed in terms of $\overline{M_{1}}$ this relation implies
absence of multifractality and it
differs from that for the pure model in~\cite{BramwellEtAl01}:

\begin{displaymath}
M_{n} = M_{1}^{n}\left(1\ + \frac{1}{(\beta J)^2}\
\frac{n(n-1)}{16 N^2}\sum_{{\bf q}\neq 0}\frac{1}{\gamma^2_{\bf
q}}\ +\ ...\right)\ ,
\end{displaymath}
since we neglected all the terms in the expansion containing
powers of $1/(\beta J)$ higher than one. In fact the relation
(\ref{moment5}) corresponds to a delta distribution of the
probability of magnetisation (when the variance is equal to zero)
that is close to the truth only for very low temperatures.

At the same time the mean value of magnetisation
which can be obtained from Eq.(\ref{moment4}) in the particular
case $p=1$:

\begin{equation}\label{Magn_Dil}
\overline{\left<m\right>} \approx c N^{-\frac{1}{4}\eta^{\dil}}\ ,
\end{equation}
recovers the finite-size scaling relation (\ref{FinSizeSc}) and accords
well with our MC simulations (see Fig.\ref{Fig1}).

\section{Conclusions}

Two important modifications, quenched site-dilution and finiteness of the
lattice, were brought to the usual 2D $XY$ model in order to investigate
quasi-long-range
ordering, which appears in this model at low temperatures, in conditions closer
to that in real many-particle systems present in nature.

We proposed a method of an analytic treatment of the 2D $XY$ model
with structural disorder based on the SWA and a sort of perturbation expansion
in the parameter characterising deviation from the pure system with
the renormalised coupling strength. Computing the perturbation expansion up to
the third order we arrived at the result for the spin-spin correlation function
decay exponent $\eta^{\dil}$, Eq.(\ref{Eq-etabetter}), which appears to be
non-universal and depends besides temperature also on concentration of non-magnetic
impurities in the system. Our analytic result shows nice accordance with MC simulations
for a wide range of dilution concentrations (Fig.\ref{Fig1}). We see that up to the
third order in the expansion the analytic results for the exponent converges to
the MC data with every next order.

We also took into account the finiteness
of the lattice which appears to be negligible for the exponent $\eta^{\dil}$
but brings on stage another important property of the 2D $XY$ model: non-vanishing
magnetisation that tends to zero with a power law as the size of the lattice
increases. Monte Carlo simulations of diluted 2D $XY$-spin systems with
different sizes, concentrations of dilution and temperatures, presented in terms
of magnetisation probability distribution, show some interesting features that
differ essentially from the case of the pure model in the same conditions.
We observed that the variance of the probability distribution function of
magnetisation, which serves as a characteristic of the distribution form, i. e.
depends on its width and flatness, is a function of temperature, system size
and concentration of dilution in contrast to the pure 2D $XY$ model where it
depends on temperature only.

We applied our analytic approach to compute the moments of magnetisation that
define the probability distribution function (and its variance) but failed to
explain the features present in our MC simulations because of the roughness
of our approximations. At the same time the analytic calculations give a
good result for the magnetisation itself in the diluted model, we found the
power law decay with system size, Eq.(\ref{Magn_Dil}), that accords with the
finite-size scaling (\ref{FinSizeSc}).

The convergence of the perturbation expansion
applied in the analytic treatment is not undoubted but from our results we can
conclude that up to the third order it gives nice accordance with the MC data.
Further analysis of this question will be a subject of a separate study.
Another direction of future
work would be to implement the same type of perturbation expansion within the
Villain model~\cite{Villain75} and to explore the deconfining
transition of the diluted model. Here the interesting part of
the question does not come across the value of the exponent $\eta$, but rather
in the way the vacancies couple to the unbinding mechanism.

\begin{acknowledgement}
{\bf Acknowledgement:} We acknowledge the CNRS-NAS exchange
programme and I. V. Stasyuk for a useful discussion.
We also thank H. Chamati for interesting correspondence.
\end{acknowledgement}

\begin{appendix}

\section*{Appendix A}
\label{appendixA}

In this appendix we give a detailed calculation of a quantity of
the type:

\begin{equation}
\overline{\left<c_{\bf r_1}\cdot\cdot\cdot c_{\bf
r_l}\cos\frac{1}{\sqrt{N}}\sum_{{\bf k}}\left( \eta^c_{\bf
k}\theta^c_{\bf k}+\eta^s_{\bf k}\theta^s_{\bf k}\right)\right>}\ ,
\end{equation}
used while computing the spin-spin
correlation function and the mean magnetisation of a finite-size
system and its higher moments.

Expanding this quantity in the parameters $\rho_{\bf
q}$'s up to the third order we have

\begin{eqnarray}\nonumber
&&\left<c_{\bf r_1}\cdot\cdot\cdot c_{\bf
r_l}\cos{\textstyle\frac{1}{\sqrt{N}}\sum_{{\bf k}}\left(
\eta^c_{\bf k}\theta^c_{\bf k}+\eta^s_{\bf k}\theta^s_{\bf
k}\right)}\right>
\\\nonumber
&&\qquad\qquad=c^l\left<\cos{\textstyle\frac{1}{\sqrt{N}}\sum_{{\bf
k}}\left( \eta^c_{\bf k}\theta^c_{\bf k}+\eta^s_{\bf
k}\theta^s_{\bf k}\right)}\right>_*\ \Bigg[\ 1
\\\nonumber
&& -\beta
{\textstyle\left(\frac{\left<H_{\rho}\cos\right>_*}{\left<\cos\right>_*}
-\left<H_{\rho}\right>_*\right)}-\beta^2{\textstyle\left(
\frac{\left<H_{\rho}\right>_*\left<H_{\rho}\cos\right>_*}
{\left<\cos\right>_*}-\left<H_{\rho}\right>^2_*\right)}
\\\nonumber
&&-\beta{\textstyle
\left(\frac{\left<H_{\rho^2}\cos\right>_*}{\left<\cos\right>_*}
-\left<H_{\rho^2}\right>_*\right) +\frac{\beta^2}{2}
\left(\frac{\left<H^2_{\rho}\cos\right>_*}{\left<\cos\right>_*}
-\left<H^2_{\rho}\right>_*\right)}
\\\nonumber
&&
+\beta^2{\textstyle\left(\frac{\left<H_{\rho}H_{\rho^2}\cos\right>_*}
{\left<\cos\right>_*}-\left<H_{\rho}H_{\rho^2}\right>_*\right)}
\\\nonumber
&& -\beta^2{\textstyle
\left(\frac{\left<H_{\rho^2}\right>_*\left<H_{\rho}\cos\right>_*}
{\left<\cos\right>_*}-\left<H_{\rho^2}\right>_*\left<H_{\rho}\right>_*\right)}
\\\nonumber
&& -\beta^2{\textstyle
\left(\frac{\left<H_{\rho}\right>_*\left<H_{\rho^2}\cos\right>_*}
{\left<\cos\right>_*}-\left<H_{\rho}\right>_*\left<H_{\rho^2}\right>_*\right)}
\\\nonumber
&&+\frac{\beta^3}{2}{\textstyle
\left(\frac{\left<H^2_{\rho}\right>_*\left<H_{\rho}\cos\right>_*}
{\left<\cos\right>_*}-\left<H^2_{\rho}\right>_*\left<H_{\rho}\right>_*\right)}
\\\nonumber
&&
+\frac{\beta^3}{2}{\textstyle\left(\frac{\left<H_{\rho}\right>_*
\left<H^2_{\rho}\cos\right>_*}
{\left<\cos\right>_*}-\left<H_{\rho}\right>_*\left<H^2_{\rho}\right>_*\right)}
\\\nonumber
&& -\beta^3{\textstyle
\left(\frac{\left<H_{\rho}\right>^2_*\left<H_{\rho}\cos\right>_*}
{\left<\cos\right>_*}-\left<H_{\rho}\right>_*^3\right)}
-\frac{\beta^3}{6}{\textstyle\left(\frac{\left<H^3_{\rho}\cos\right>_*}
{\left<\cos\right>_*}-\left<H_{\rho}^3\right>_*\right)}
\\\nonumber
&& +\frac{\beta}{c} \sum_{\bf q}\sum_{i=1}^le^{i{\bf
qr}_i}\rho_{\bf q}\Big[\ {\textstyle
\left(\frac{\left<H_{\rho}\cos\right>_*}{\left<\cos\right>_*}
-\left<H_{\rho}\right>_*\right)}
\\\nonumber && + {\textstyle\left(\frac{\left<H_{\rho^2}\cos\right>_*}{\left<\cos\right>_*}
-\left<H_{\rho^2}\right>_*\right)}
-\frac{\beta}{2}
{\textstyle\left(\frac{\left<H^2_{\rho}\cos\right>_*}{\left<\cos\right>_*}
-\left<H^2_{\rho}\right>_*\right)}
\\\nonumber
&& +\ \beta {\textstyle\left(\frac{\left<H_{\rho}\right>_*\left<H_{\rho}\cos\right>_*}
{\left<\cos\right>_*}-\left<H_{\rho}\right>^2_*\right)}\ \Big]
\\\nonumber
&&-\frac{\beta}{c^2}\sum_{\bf q,
q'}\sum_{i=1}^l\sum_{j=1}^le^{i({\bf qr}_i+{\bf q'r}_j)}\rho_{\bf
q}\rho_{\bf q'}{\textstyle\left(\frac{\left<H_{\rho}\cos\right>_*}
{\left<\cos\right>_*} -\left<H_{\rho}\right>_*\right)}
\\\nonumber
&& -\frac{1}{c}\sum_{\bf q}\sum_{i=1}^le^{i{\bf qr}_i}\rho_{\bf q}
+\frac{1}{c^2}\sum_{\bf q}\sum_{\bf
q'}\sum_{i=1}^l\sum_{j=1}^le^{i({\bf qr}_i+{\bf q'r}_j)}\rho_{\bf
q}\rho_{\bf q'}
\\
&& +\frac{1}{c^3} \sum_{\bf
q,q',q''}\sum_{i=1}^l\sum_{j=1}^l\sum_{k=1}^le^{i({\bf qr}_i+{\bf
q'r}_j+{\bf q''r}_k)}\rho_{\bf q}\rho_{\bf q'}\rho_{\bf q''}\ \Bigg]
\label{monster}
\end{eqnarray}
where in the brackets we have noted for economy of space \newline
$\cos\frac{1}{\sqrt{N}}\sum_{\bf k}\left(\eta^c_{\bf
k}\theta^c_{\bf k}+\eta^s_{\bf k}\theta^s_{\bf
k}\right)\equiv\cos$ . One can easily find

\begin{eqnarray}\label{CosMean}
&&\left<\cos{\textstyle\frac{1}{\sqrt{N}}\sum_{{\bf k}}\left(
\eta^c_{\bf k}\theta^c_{\bf k}+\eta^s_{\bf k}\theta^s_{\bf
k}\right)}\right>_*
\\\nonumber
&&\quad=\ \Re\left<e^{\frac{i}{\sqrt{N}}\sum_{{\bf k}}\left(
\eta^c_{\bf k}\theta^c_{\bf k}+\eta^s_{\bf k}\theta^s_{\bf
k}\right)}\right>_* =\ e^{-\frac{1}{4c^2\beta JN}\sum_{{\bf k}\neq
0}\frac{\eta_{\bf k}\eta_{\bf -k}}{\gamma_{\bf k}}}\ ,\nonumber
\end{eqnarray}
where $\eta_{\bf k}\equiv\eta^c_{\bf k}+i\eta^s_{\bf k}$ and the
sum runs over the whole 1st Brillouin zone except the point $k=0$.
The quantities of the form $\left<\theta_{\bf k_1} \cdot\cdot\cdot
\theta_{\bf k_n}\cos\frac{1}{\sqrt{N}}\sum_{{\bf k}}\left(
\eta^c_{\bf k}\theta^c_{\bf k}+\eta^s_{\bf k}\theta^s_{\bf
k}\right)\right>_*$ in (\ref{monster}) can be obtained using the
property:

\begin{eqnarray}
&&\left<\theta_{\bf k_1} \cdot\cdot\cdot \theta_{\bf
k_n}\cos{\textstyle\frac{1}{\sqrt{N}}\sum_{\bf k}\left(
\eta^c_{\bf k}\theta^c_{\bf k}+\eta^s_{\bf k}\theta^s_{\bf
k}\right)}\right>
\\\nonumber
&&= {\textstyle\left(\frac{\sqrt{N}}{2i}\right)^n}
\frac{\partial}{\partial\eta_{\bf k_1}} \cdot\cdot\cdot
\frac{\partial}{\partial\eta_{\bf
k_n}}\left<\cos{\textstyle\frac{1}{\sqrt{N}}\sum_{\bf k}\left(
\eta^c_{\bf k}\theta^c_{\bf k}+\eta^s_{\bf k}\theta^s_{\bf
k}\right)}\right>,
\end{eqnarray}
where the notation $\frac{\partial}{\partial\eta_{\bf
k}}\equiv\frac{\partial}{\partial\eta^c_{\bf
k}}+i\frac{\partial}{\partial\eta^s_{\bf k}}$ is used.

Now, taking derivatives step by step and applying the obvious
equality $\frac{\partial\eta_{\bf k}}{\partial\eta_{\bf k'}} = 2
\delta_{{\bf k+k'},0}$, we obtain:

\begin{eqnarray}\label{theta2}
&&\left<\theta_{\bf k}\theta_{\bf
k'}\cos{\textstyle\frac{1}{\sqrt{N}}\sum_{{\bf k}}\left(
\eta^c_{\bf k}\theta^c_{\bf k}+\eta^s_{\bf k}\theta^s_{\bf
k}\right)}\right>_*
\\\nonumber
&&\qquad=\ \left<\cos{\textstyle\frac{1}{\sqrt{N}}\sum_{{\bf
k}}\left( \eta^c_{\bf k}\theta^c_{\bf k}+\eta^s_{\bf
k}\theta^s_{\bf k}\right)}\right>_*
\\\nonumber
&&\qquad\times \left[\ -\ \frac{1}{(2c^2\beta
J)^2N}\frac{\eta_{\bf k}\eta_{\bf k'}}{\gamma_{\bf k}\gamma_{\bf
k'}}\ + \ \frac{1}{2c^2\beta J}\frac{\delta_{{\bf
k+k'},0}}{\gamma_{\bf k}}\right]\ \ ,
\end{eqnarray}

\begin{eqnarray}\label{theta4}
&&\left<\theta_{\bf k_1}\theta_{\bf k_2}\theta_{\bf
k_3}\theta_{\bf k_4}\cos{\textstyle\frac{1}{\sqrt{N}}\sum_{{\bf
k}}\left( \eta^c_{\bf k}\theta^c_{\bf k}+\eta^s_{\bf
k}\theta^s_{\bf k}\right)}\right>_*\
\\\nonumber
&&=\ \left<\cos{\textstyle\frac{1}{\sqrt{N}}\sum_{{\bf k}}\left(
\eta^c_{\bf k}\theta^c_{\bf k}+\eta^s_{\bf k}\theta^s_{\bf
k}\right)}\right>_*
\\\nonumber
&&\times\Bigg[\ \frac{1}{(2c^2\beta J)^4N^2}\ \frac{\eta_{\bf
k_1}\eta_{\bf k_2}\eta_{\bf k_3}\eta_{\bf k_4}}{\gamma_{\bf
k_1}\gamma_{\bf k_2}\gamma_{\bf k_3}\gamma_{\bf k_4}}
\\\nonumber
&&-\ \frac{1}{(2c^2\beta J)^3N}\Bigg(\frac{\delta_{{\bf
k_1+k_2},0}\eta_{\bf k_3}\eta_{\bf k_4}}{\gamma_{\bf
k_2}\gamma_{\bf k_3}\gamma_{\bf k_4}} + \frac{\delta_{{\bf
k_1+k_3},0}\eta_{\bf k_2}\eta_{\bf k_4}}{\gamma_{\bf
k_2}\gamma_{\bf k_3}\gamma_{\bf k_4}}
\\\nonumber
&&\qquad\qquad\qquad\ + \frac{\delta_{{\bf k_1+k_4},0}\eta_{\bf
k_2}\eta_{\bf k_3}}{\gamma_{\bf k_2}\gamma_{\bf k_3}\gamma_{\bf
k_4}}
+ \frac{\delta_{{\bf k_2+k_3},0}\eta_{\bf k_1}\eta_{\bf
k_4}}{\gamma_{\bf k_1}\gamma_{\bf k_2}\gamma_{\bf k_4}}
\\\nonumber
&&\qquad\qquad\qquad\ + \frac{\delta_{{\bf k_2+k_4},0}\eta_{\bf
k_1}\eta_{\bf k_3}}{\gamma_{\bf k_1}\gamma_{\bf k_2}\gamma_{\bf
k_3}} + \frac{\delta_{{\bf k_3+k_4},0}\eta_{\bf k_1}\eta_{\bf
k_2}}{\gamma_{\bf k_1}\gamma_{\bf k_2}\gamma_{\bf k_3}}\Bigg)
\\\nonumber
&&\quad\ + \frac{1}{(2c^2\beta J)^2}\Bigg(\frac{\delta_{{\bf
k_1+k_2},0}\delta_{{\bf k_3+k_4},0}}{\gamma_{\bf k_1}\gamma_{\bf
k_2}} + \frac{\delta_{{\bf k_1+k_3},0}\delta_{{\bf
k_2+k_4},0}}{\gamma_{\bf k_1}\gamma_{\bf k_2}}
\\\nonumber
&&\qquad\qquad\qquad\qquad\qquad\qquad\qquad + \frac{\delta_{{\bf
k_1+k_4},0}\delta_{{\bf k_2+k_3},0}}{\gamma_{\bf k_1}\gamma_{\bf
k_2}} \Bigg)\Bigg]
\end{eqnarray}
and

\begin{eqnarray}\label{theta6}
&&\left<\theta_{\bf k_1}\theta_{\bf k_2}\theta_{\bf
k_3}\theta_{\bf k_4}\theta_{\bf k_5}\theta_{\bf
k_6}\cos{\textstyle\frac{1}{\sqrt{N}}\sum_{{\bf k}}\left(
\eta^c_{\bf k}\theta^c_{\bf k}+\eta^s_{\bf k}\theta^s_{\bf
k}\right)}\right>_*\
\\\nonumber
&&=\ \left<\cos{\textstyle\frac{1}{\sqrt{N}}\sum_{{\bf k}}\left(
\eta^c_{\bf k}\theta^c_{\bf k}+\eta^s_{\bf k}\theta^s_{\bf
k}\right)}\right>_*
\\\nonumber
&& \times\Bigg[\ -\frac{1}{(2c^2\beta J)^6N^3}\ \frac{\eta_{\bf
k_1}\eta_{\bf k_2}\eta_{\bf k_3}\eta_{\bf k_4}\eta_{\bf
k_5}\eta_{\bf k_6}}{\gamma_{\bf k_1}\gamma_{\bf k_2}\gamma_{\bf
k_3}\gamma_{\bf k_4}\gamma_{\bf k_5}\gamma_{\bf k_6}}
\\\nonumber
&&+\ \frac{1}{(2c^2\beta J)^5N^2}\Bigg(\ \frac{\delta_{{\bf
k_1+k_2},0}\eta_{\bf k_3}\eta_{\bf k_4}\eta_{\bf k_5}\eta_{\bf
k_6}}{\gamma_{\bf k_2}\gamma_{\bf k_3}\gamma_{\bf k_4}\gamma_{\bf
k_5}\gamma_{\bf k_6}}\ +\ \cdot\ \cdot\ \cdot\ \Bigg)
\\\nonumber
&&-\ \frac{1}{(2c^2\beta J)^4N}\Bigg(\ \frac{\delta_{{\bf
k_1+k_2},0}\delta_{{\bf k_3+k_4},0}\eta_{\bf k_5}\eta_{\bf
k_6}}{\gamma_{\bf k_1}\gamma_{\bf k_3}\gamma_{\bf k_5}\gamma_{\bf
k_6}}\ + \ \cdot\ \cdot\ \cdot\ \Bigg)
\\\nonumber
&&+\ \frac{1}{(2c^2\beta J)^3}\Bigg(\ \frac{\delta_{{\bf
k_1+k_2},0}\delta_{{\bf k_3+k_4},0}\delta_{{\bf
k_5+k_6},0}}{\gamma_{\bf k_1}\gamma_{\bf k_3}\gamma_{\bf k_5}}\ +
\ \cdot\ \cdot\ \cdot\ \Bigg)\ \Bigg]\ ,\nonumber
\end{eqnarray}
where the sums in the brackets run over all possible ways of
choosing the pairs of ${\bf k}$'s in the delta-symbols. Averages
$\left<\theta_{\bf k}\theta_{\bf k'}\right>_*$, $\left<\theta_{\bf
k_1}\theta_{\bf k_2}\theta_{\bf k_3}\theta_{\bf k_4}\right>_*$ and
$\left<\theta_{\bf k_1}\theta_{\bf k_2}\theta_{\bf k_3}\theta_{\bf
k_4}\theta_{\bf k_5}\theta_{\bf k_6}\right>_*$ can be obtained
from (\ref{theta2})-(\ref{theta6}) putting all the $\eta_{\bf
k}$'s equal to zero.

Substituting these results in (\ref{monster}) and applying
(\ref{Avrg}) we arrive at the expression:

\begin{eqnarray}\label{appAresult}
&&\overline{\left<c_{\bf r_1}\cdot\cdot\cdot c_{\bf
r_l}\cos{\textstyle\frac{1}{\sqrt{N}}\sum_{\bf k}\left(
\eta^c_{\bf k}\theta^c_{\bf k}+\eta^s_{\bf k}\theta^s_{\bf
k}\right)}\right>}
\\\nonumber
&&= c^l \left<\cos{\textstyle\frac{1}{\sqrt{N}}\sum_{\bf k}\left(
\eta^c_{\bf k}\theta^c_{\bf k}+\eta^s_{\bf k}\theta^s_{\bf
k}\right)}\right>_*\ \Bigg[\ 1\ -\ \frac{1}{\beta J}
\\\nonumber &&
\times\ \Bigg(\frac{1-c}{4c^3}\frac{1}{N^2}\sum_{{\bf k},{\bf
k'}}g_{\bf k, k'}g_{\bf k', k}\frac{\eta_{\bf k}\eta_{\bf
-k}}{\gamma_{\bf k}}\ - \frac{1-3c+2c^2}{2c^4}
\\\nonumber
&& \times\Big(\frac{1}{N}\sum_{{\bf k}}{\textstyle\frac{\eta_{\bf
k}\eta_{\bf -k}}{\gamma_{\bf k}}} - \frac{1}{2N^3}\sum_{{\bf
k},{\bf k'},{\bf k''}}g_{\bf k, k'}g_{\bf k', k''}g_{\bf k'',
k}{\textstyle\frac{\eta_{\bf k}\eta_{\bf -k}}{\gamma_{\bf
k}}}\Big)\Bigg)\Bigg]
\end{eqnarray}
with $g_{\bf k, k'}$ given by (\ref{def_of_g}). We have neglected
terms that vanish in the thermodynamic limit and terms containing
higher powers of $1/(\beta J)$, since we consider low
temperatures.

\section*{Appendix B}
\label{appendixB}

We are interested in the asymptotic behaviour of the sums:

\begin{eqnarray}\nonumber
S_1(R,N)&\equiv&\frac{1}{N}\sum_{{\bf q}\neq
0}{\textstyle\frac{\sin^2\frac{\bf qR}{2}}{\gamma_{\bf q}}}\ ,
\\\nonumber
S_2(R,N)&\equiv&\frac{1}{N^2}\sum_{\bf q, q'}g_{\bf q, q'}g_{\bf
q', q}{\textstyle\frac{\sin^2\frac{\bf qR}{2}}{\gamma_{\bf
q}}}\ ,\label{AppB1}
\\\nonumber
S_3(R,N)&\equiv&\frac{1}{N^3}\sum_{\bf q, q', q''}g_{\bf -q,
q'}g_{\bf q', q''}g_{\bf q'', q}{\textstyle\frac{\sin^2\frac{\bf
qR}{2}}{\gamma_{\bf q}}}\ ,
\\\nonumber
\tilde{S}_1(N)&\equiv&\frac{1}{N}\sum_{{\bf q}\neq
0}{\textstyle\frac{1}{\gamma_{\bf q}}}\ ,
\\\nonumber
\tilde{S}_2( N)&\equiv&\frac{1}{N^2}\sum_{\bf q, q'}g_{\bf q,
q'}g_{\bf q', q}{\textstyle\frac{1}{\gamma_{\bf q}}}\ ,
\\\nonumber
\tilde{S}_3(N)&\equiv&\frac{1}{N^3}\sum_{\bf q, q', q''}g_{\bf -q,
q'}g_{\bf q', q''}g_{\bf q'', q}{\textstyle\frac{1}{\gamma_{\bf
q}}}\ ,\label{AppB2}
\end{eqnarray}
with $g_{\bf q, q'}$ given by Eq.(\ref{def_of_g}), when
$R\rightarrow\infty$ and $N\rightarrow\infty$.

The singularity in $\frac{1}{\gamma_{\bf q}}$ in the point $q=0$
defines the asymptotic behaviour of the sums. Thus we will have
the same asymptotic behaviour after expanding the expressions
in the sums for small ${\bf q}$'s:

\begin{displaymath}
S_1(R,N)\ =\ c_1\ +\ \frac{2}{a^2}\ \frac{1}{N}\sum_{{\bf q}\neq
0}\frac{\sin^2{\textstyle\frac{\bf qR}{2}}}{|{\bf q}|}\
,\quad\phantom{*******}
\end{displaymath}

\begin{eqnarray}\nonumber
S_2(R,N)\ =\ c_2 &+& \frac{1}{2N}\sum_{{\bf k}\neq 0}\left(g_{\bf k,
k}-g_{\bf k, -k}\right)
\\\nonumber
&\times&\frac{2}{a^2}\frac{1}{N}\sum_{{\bf q}\neq
0}\frac{\sin^2{\textstyle\frac{\bf qR}{2}}}{|{\bf q}|}\
,\phantom{*********}
\end{eqnarray}

\begin{eqnarray}\nonumber
S_3(R,N)\ =\ c_3 &-& \frac{1}{2N}\sum_{{\bf k}\neq 0}\sum_{{\bf
k'}\neq 0}g_{\bf k, k'}\left(g_{\bf k, k'}-g_{\bf k, -k'}\right)
\\\nonumber
&\times&\frac{2}{a^2}\frac{1}{N}\sum_{{\bf q}\neq
0}\frac{\sin^2{\textstyle\frac{\bf qR}{2}}}{|{\bf q}|}\
,
\end{eqnarray}

\begin{displaymath}
\tilde{S}_1(N)\ =\ \tilde{c}_1\ +\ \frac{2}{a^2}\
\frac{1}{N}\sum_{{\bf q}\neq 0}\frac{1}{|{\bf q}|}\
,\quad\phantom{*******}
\end{displaymath}

\begin{eqnarray}\nonumber
\tilde{S}_2(N)\ =\ \tilde{c}_2 &+& \frac{1}{2N}\sum_{{\bf k}\neq
0}\left(g_{\bf k, k}-g_{\bf k, -k}\right)
\\\nonumber
&\times&\frac{2}{a^2}\frac{1}{N}\sum_{{\bf q}\neq 0}\frac{1}{|{\bf
q}|}\ ,\phantom{*********}
\end{eqnarray}

\begin{eqnarray}\nonumber
\tilde{S}_3(N)\ =\ \tilde{c}_3 &-& \frac{1}{2N}\sum_{{\bf k}\neq
0}\sum_{{\bf k'}\neq 0}g_{\bf k, k'}\left(g_{\bf k, k'}-g_{\bf k,
-k'}\right)
\\\nonumber
&\times&\frac{2}{a^2}\frac{1}{N}\sum_{{\bf q}\neq 0}\frac{1}{|{\bf
q}|}\ .
\end{eqnarray}
$c_1$, $c_2$, $c_3$, $\tilde{c}_1$, $\tilde{c}_2$, $\tilde{c}_3$ are
constants.

Numerical calculation gives

\begin{eqnarray}\nonumber
&&\frac{1}{2N}\sum_{{\bf k}\neq 0}\left(g_{\bf k, k}-g_{\bf k,
-k}\right)\ \approx\ 0.73\ ,
\\\nonumber
&&\frac{1}{2N}\sum_{{\bf k}\neq 0}\sum_{{\bf k'}\neq 0}g_{\bf k,
k'}\left(g_{\bf k, k'}-g_{\bf k, -k'}\right)\ \approx\ 0.27\ .
\end{eqnarray}

To get the asymptotic behaviour of the sums

\begin{displaymath}
\frac{2}{a^2} \frac{1}{N}\sum_{{\bf q}\neq
0}\frac{\sin^2{\textstyle\frac{\bf qR}{2}}}{|{\bf q}|}\qquad ,
\frac{2}{a^2} \frac{1}{N}\sum_{{\bf q}\neq 0}\frac{1}{|{\bf q}|}
\end{displaymath}
we replace sums over the 1st Brillouin zone in the thermodynamic
limit with integrals over continuous variables $q_x$, $q_y$,
according to the formula

\begin{displaymath}
\sum_{\bf q}\ \rightarrow\ \frac{Na^2}{(2\pi)^2}\int dq_x\int dq_y
+ o(N^{-1})
\end{displaymath}
We take into account the absence of the terms with $q=0$ in the
sums cutting out from the continuous domains of integration spaces
around the points $q=0$ with area equal to
$\frac{(2\pi)^2}{Na^2}$:

\begin{eqnarray}\nonumber
\left\{ \begin{array}{ll} q_x\ \epsilon\
\left(-\frac{\pi}{a},-\frac{\pi}{a\sqrt{N}}\right) \cup
\left(\frac{\pi}{a\sqrt{N}},\frac{\pi}{a}\right);\\
q_y\ \epsilon\ \left(-\frac{\pi}{a},-\frac{\pi}{a\sqrt{N}}\right)
\cup \left(\frac{\pi}{a\sqrt{N}},\frac{\pi}{a}\right).
\end{array}\right.
\end{eqnarray}

\begin{figure}
\setlength{\unitlength}{1cm}
\begin{picture}(10,4)

\newsavebox{\hop}
\savebox{\hop}(5,4)[bl] {\put(2,0){\line(0,1){4}}
\put(3,0){\line(0,1){4}} \put(4,0){\line(0,1){4}}
\put(1,1){\line(1,0){4}} \put(1,2){\line(1,0){4}}
\put(1,3){\line(1,0){4}} \put(2.5,1.5){\line(1,0){1}}
\put(2.5,1.5){\line(0,1){1}} \put(3.5,2.5){\line(-1,0){1}}
\put(3.5,2.5){\line(0,-1){1}} \put(2,1){\circle*{.2}}
\put(3,2){\circle*{.2}} \put(2,2){\circle*{.2}}
\put(2,3){\circle*{.2}} \put(3,1){\circle*{.2}}
\put(4,1){\circle*{.2}} \put(3,3){\circle*{.2}}
\put(4,2){\circle*{.2}} \put(4,3){\circle*{.2}}
\put(5,1.4){\vector(-4,1){1.8}}
}

\newsavebox{\hopp}
\savebox{\hopp}(5,4)[bl] {\put(2,0){\line(0,1){4}}
\put(3,0){\line(0,1){4}} \put(4,0){\line(0,1){4}}
\put(1,1){\line(1,0){3.8}} \put(1,2){\line(1,0){3.8}}
\put(1,3){\line(1,0){3.8}} \put(3,2){\circle{1.2}}
\put(2,1){\circle*{.2}} \put(3,2){\circle*{.2}}
\put(2,2){\circle*{.2}} \put(2,3){\circle*{.2}}
\put(3,1){\circle*{.2}} \put(4,1){\circle*{.2}}
\put(3,3){\circle*{.2}} \put(4,2){\circle*{.2}}
\put(4,3){\circle*{.2}} \put(1,1.4){\vector(4,1){1.8}}
\put(0,1.2){${\bf q}=0$} }

\newsavebox{\hoppp}
\savebox{\hoppp}(4,3)[bl]{\put(1,1){\line(0,1){1}}
\put(1,1){\line(1,0){1}} \put(2,2){\line(-1,0){1}}
\put(2,2){\line(0,-1){1}} \put(2.5,1.5){$=$}
\put(3.5,1.5){\circle{1.2}} \put(1.5,1.5){$S$} \put(3.5,1.5){$S$}
}

\put(-1,0){\usebox{\hop}} \put(4,0){\usebox{\hopp}}

\end{picture}
\caption{The cut out of the point ${\bf q}=0$}
\end{figure}
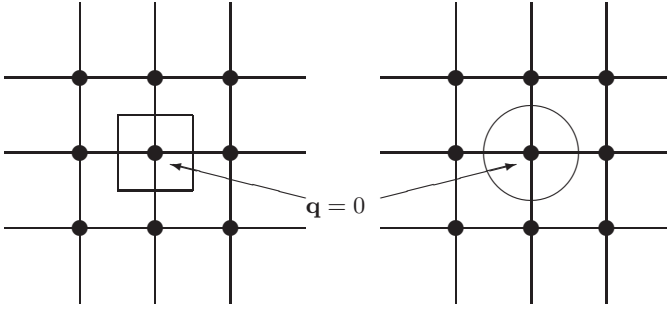

\vspace{0.5cm} We should stress here that the exact value of this
area is not important for the asymptotic behaviour which is the
point of our interest, it only must be proportional to $1/N$.

After passing to polar coordinates $q\ \epsilon\
\left(\frac{2\sqrt{\pi}}{a\sqrt{N}},\frac{2 \sqrt{\pi}}{a}\right)$
and $\varphi\ \epsilon\ (0,2\pi)$ one can write

\begin{equation}
\frac{2}{a^2} \frac{1}{N}\sum_{{\bf q}\neq
0}\frac{\sin^2{\textstyle\frac{\bf qR}{2}}}{|{\bf q}|} =
\frac{1}{2\pi^2}\int_{\frac{2\sqrt{\pi}}{a\sqrt{N}}}
^{\frac{2\sqrt{\pi}}{a}}dq\int_{0}^{2\pi}d\varphi
\frac{\sin^2\frac{qR\cos\varphi}{2}}{q} \label{AppB7}
\end{equation}
and

\begin{equation}
\frac{2}{a^2} \frac{1}{N}\sum_{{\bf q}\neq 0}\frac{1}{|{\bf q}|} =
\frac{1}{2\pi^2}\int_{\frac{2\sqrt{\pi}}{a\sqrt{N}}}
^{\frac{2\sqrt{\pi}}{a}}\frac{dq}{q}\int_{0}^{2\pi}d\varphi\
.\label{AppB9}
\end{equation}

The integral in (\ref{AppB9}) gives $\frac{1}{2\pi}\ln N$. So

\begin{eqnarray}\nonumber
&&\tilde{S}_1(N)\ =\ \tilde{c}_1\ +\ \frac{1}{2\pi}\ln N\ ,
\\\nonumber
&&\tilde{S}_2(N)\ =\ \tilde{c}_2+ 0.73\frac{1}{2\pi}\ln N\ ,
\\\nonumber
&&\tilde{S}_3(N)\ =\ \tilde{c}_3- 0.27\frac{1}{2\pi}\ln N\ .
\end{eqnarray}

Now we are interested in the behaviour of the integral in
(\ref{AppB7}) in the asymptotic case: $N\rightarrow\infty$,
$R\rightarrow\infty$ . After change of variables,
$\frac{kR}{2}\rightarrow x$, we split the integrals over $x$ in two
parts:

\begin{displaymath}
\int_{\frac{R\sqrt{\pi}}{a\sqrt{N}}}
^{\frac{R\sqrt{\pi}}{a}}dx\quad \rightarrow\quad
\int_{\frac{R\sqrt{\pi}}{a\sqrt{N}}} ^{\varepsilon}dx\ +\
\int_{\varepsilon} ^{\frac{R\sqrt{\pi}}{a}}dx\ .
\end{displaymath}
It is reasonable to assume that the ratio $\frac{(R/a)}{\sqrt{N}}$ is
small in the thermodynamical limit, then we choose $\varepsilon$
small enough to change $\sin(x\cos\varphi)$ in the limits
$\left(\frac{R\sqrt{\pi}}{a\sqrt{N}},\varepsilon\right)$ by its
argument. In the integral over the rest of the domain,
$\left(\varepsilon,\frac{R\sqrt{\pi}}{a}\right)$, we substitute
$\sin^2$ with its mean value $1/2$, this cannot
change the asymptotic behaviour for a large $R$. Thus we have simply
integrable functions now and one easily finds:

\begin{eqnarray}\nonumber
&&\frac{2}{a^2} \frac{1}{N}\sum_{{\bf q}\neq
0}\frac{\sin^2{\textstyle\frac{\bf qR}{2}}}{|{\bf q}|}\ =\
\frac{1}{2\pi^2}\int_{\frac{R\sqrt{\pi}}{a\sqrt{N}}}
^{\varepsilon}xdx\int_{0}^{2\pi}d\varphi\cos^2\varphi\
\\
&& +\ \frac{1}{4\pi^2}\int_{\varepsilon}
^{\frac{R\sqrt{\pi}}{a}}\frac{dx}{x}\int_{0}^{2\pi}d\varphi\ =\ C\
+\ \frac{1}{2\pi}\ln\frac{R}{a}\ .\qquad\quad\phantom{}\label{AppB12}
\end{eqnarray}
We neglected here terms containing small values $(R/a)^2/N$ and
$1/N$ and collected terms with the fixed parameter $\varepsilon$
in the constant $C$.

Finally we have:

\begin{eqnarray}\nonumber
&&S_1(R,N)\ =\ c_1'\ +\ \frac{1}{2\pi}\ln \textstyle{\frac{R}{a}}\ ,
\\\nonumber
&&S_2(R,N)\ =\ c_2'+ 0.73\frac{1}{2\pi}\ln \textstyle{\frac{R}{a}}\
,
\\\nonumber
&&S_3(R,N)\ =\ c_3'- 0.27\frac{1}{2\pi}\ln \textstyle{\frac{R}{a}}\ .
\end{eqnarray}
Note, that the above estimates hold for finite $N$ as well.

\end{appendix}

%
%
%
%
%

\end{document}